\documentclass[longauth]{aa}

\usepackage{graphicx}
\graphicspath{{./Input/}}
\usepackage{txfonts}

\usepackage{amssymb}
\usepackage{latexsym}
\usepackage{amstext}
\usepackage{astro_bib_macro}
\usepackage{natbib}
\bibpunct{(}{)}{;}{a}{}{,}

\newcommand{\unldots}{..}
\newcommand{\deuxvdots}{:}

\usepackage{color}

\begin{document}
\title{Interferometric data reduction with AMBER/VLTI.\\ Principle,
  estimators and illustration.}
%
%
%
\author{%
       E.~Tatulli\inst{1,2}
  \and F.~Millour\inst{1,3}
  \and A.~Chelli\inst{1}
  \and G.~Duvert\inst{1}
  \and B.~Acke\inst{1,12}
  \and K.-H.~Hofmann\inst{4}
  \and S.~Kraus\inst{4}
  \and F.~Malbet\inst{1}
  \and P.~M\`ege\inst{1}
  \and R.G.~Petrov\inst{3}
  \and O.~Hernandez-Utrera\inst{3}
  \and M.~Vannier\inst{3,13}
  \and G.~Zins\inst{1}
  \and P.~Antonelli\inst{5}
  \and U.~Beckmann\inst{4}
  \and Y.~Bresson\inst{5}
  \and L.~Gl\"uck\inst{1}
  \and P.~Kern\inst{1}
  \and S.~Lagarde\inst{5}
  \and E.~Le~Coarer\inst{1}
  \and F.~Lisi\inst{2}
  \and K.~Perraut\inst{1}
  \and S.~Robbe-Dubois\inst{3}
  \and A.~Roussel\inst{5}
  \and M.~Accardo\inst{2}
  \and K.~Agabi\inst{3}
  \and B.~Arezki\inst{1}
  \and E.~Aristidi\inst{3}
  \and C.~Baffa\inst{2}
  \and J.~Behrend\inst{4}
  \and T.~Bl\"ocker\inst{4}
  \and S.~Bonhomme\inst{5}
  \and S.~Busoni\inst{2}
  \and F.~Cassaing\inst{6}
  \and J.-M.~Clausse\inst{5}
  \and J.~Colin\inst{5}
  \and C.~Connot\inst{4}
  \and A.~Delboulb\'e\inst{1}
  \and T.~Driebe\inst{4}
  \and M.~Dugu\'e\inst{5}
  \and P.~Feautrier\inst{1}
  \and D.~Ferruzzi\inst{2}
  \and T.~Forveille\inst{1}
  \and E.~Fossat\inst{3}
  \and R.~Foy\inst{7}
  \and D.~Fraix-Burnet\inst{1}
  \and A.~Gallardo\inst{1}
  \and S.~Gennari\inst{2}
  \and A.~Glentzlin\inst{5}
  \and E.~Giani\inst{2}
  \and C.~Gil\inst{1,14}
  \and M.~Heiden\inst{4}
  \and M.~Heininger\inst{4}
  \and D.~Kamm\inst{5}
  \and D.~Le Contel\inst{5}
  \and J.-M.~Le Contel\inst{5}
  \and B.~Lopez\inst{5}
  \and Y.~Magnard\inst{1}
  \and A.~Marconi\inst{2}
  \and G.~Mars\inst{5}
  \and G.~Martinot-Lagarde\inst{8,15}
  \and P.~Mathias\inst{5}
  \and J.-L.~Monin\inst{1}
  \and D.~Mouillet\inst{1,16}
  \and D.~Mourard\inst{5}
  \and E.~Nussbaum\inst{4}
  \and K.~Ohnaka\inst{4}
  \and J.~Pacheco\inst{5}
  \and F.~Pacini\inst{2}
  \and C.~Perrier\inst{1}
  \and P.~Puget\inst{1}
  \and Y.~Rabbia\inst{5}
  \and S.~Rebattu\inst{5}
  \and F.~Reynaud\inst{9}
  \and A.~Richichi\inst{10}
  \and M.~Sacchettini\inst{1}
  \and P.~Salinari\inst{2}
  \and D.~Schertl\inst{4}
  \and W.~Solscheid\inst{4}
  \and T.~Preibisch\inst{4}
  \and P.~Stee\inst{5}
  \and P.~Stefanini\inst{2}
  \and M.~Tallon\inst{7}
  \and I.~Tallon-Bosc\inst{7}
  \and D.~Tasso\inst{5}
  \and L.~Testi\inst{2}
  \and J.-C.~Valtier\inst{5}
  \and N.~Ventura\inst{1}
  \and O.~Von der L\"uhe\inst{11} 	
  \and G.~Weigelt\inst{4}
%
%
}             
 
\offprints{E.~Tatulli\\ email: \texttt{<etatulli@arcetri.astro.it>}}
\institute{
  Laboratoire d'Astrophysique de Grenoble, UMR 5571 Universit\'e Joseph
  Fourier/CNRS, BP 53, F-38041 Grenoble Cedex 9, France
  \and INAF-Osservatorio Astrofisico di Arcetri, Istituto Nazionale di
  Astrofisica, Largo E. Fermi 5, I-50125 Firenze, Italy
  \and Laboratoire Universitaire d'Astrophysique de Nice, UMR 6525
  Universit\'e de Nice/CNRS, Parc Valrose, F-06108 Nice cedex 2, France
  \and Max-Planck-Institut f\"ur Radioastronomie, Auf dem H\"ugel 69,
  D-53121 Bonn, Germany
  \and Laboratoire Gemini, UMR 6203 Observatoire de la C\^ote
  d'Azur/CNRS, BP 4229, F-06304 Nice Cedex 4, France
  \and ONERA/DOTA, 29 av de la Division Leclerc, BP 72, F-92322
  Chatillon Cedex, France 
  \and Centre de Recherche Astronomique de Lyon, UMR 5574 Universit\'e
  Claude Bernard/CNRS, 9 avenue Charles Andr\'e, F-69561 Saint Genis
  Laval cedex, France
  \and Division Technique INSU/CNRS UPS 855, 1 place Aristide
  Briand, F-92195 Meudon cedex, France
  \and IRCOM, UMR 6615 Universit\'e de Limoges/CNRS, 123 avenue Albert
  Thomas, F-87060 Limoges cedex, France
  \and European Southern Observatory, Karl Schwarzschild Strasse 2,
  D-85748 Garching, Germany
  \and Kiepenheuer-Institut f\"ur Sonnenphysik, Schoeneckstr. 6-7, 79104 Freiburg, Germany
  \and Instituut voor Sterrenkunde, KULeuven, Celestijnenlaan 200B,
  B-3001 Leuven, Belgium 
  \and European Southern Observatory, Casilla 19001, Santiago 19,
  Chile
  \and Centro de Astrofisica da Universidade do Porto, Rua das Estrelas,
4150-762 Porto, Portugal
  \and \emph{Present affiliation:} Observatoire de la C\^ote d'Azur -
  Calern, 2130 Route de l'Observatoire , F-06460 Caussols, France
  \and \emph{Present affiliation:} Laboratoire Astrophysique de
  Toulouse, UMR 5572 Universit\'e Paul Sabatier/CNRS, BP 826, F-65008
  Tarbes cedex, France 
}

\date{Received date; accepted date}
\abstract{}{We present in this paper an innovative data reduction method for single-mode interferometry. It has been specifically developed for the AMBER instrument, the three-beam combiner of the Very Large Telescope Interferometer, but can be derived for any single-mode interferometer.}{The algorithm is based on a direct modelling of the fringes in the detector plane. As such, it requires a preliminary calibration of the instrument in order to obtain the calibration matrix which builds the linear relationship between the interferogram and the interferometric observable, that is the complex visibility. Once the calibration procedure has been performed, the  signal processing appears to be a classical least square determination of a linear inverse problem. From the estimated complex visibility, we derive the squared visibility, the closure phase and the spectral differential phase.}{The data reduction procedures are gathered into the so-called amdlib software, now available for the community, and presented in this paper. Furthermore, each step of this original algorithm is illustrated and discussed from various on-sky observations conducted with the VLTI, with a focus on the control of the data quality and the effective execution of the data reduction procedures. We point out the present limited performances of the instrument due to VLTI instrumental vibrations, difficult to calibrate.}{}%
  \keywords{ %
    Technique: interferometric -- methods: data analysis -- instrumentation: interferometers
    }
\maketitle
%
\section{Introduction}

AMBER is the first-generation near-infrared three-way beam combiner
\citep{petrov_1} of the \emph{Very Large Telescope Interferometer
  (VLTI)}. This instrument provides simultaneously spectrally
dispersed visibility for
three baselines and a closure phase at three different spectral
resolution. AMBER has been designed to investigate the milli-arcsec
surrounding of astrophysical sources like young and evolved stars,
active galactic nuclei and possibly detect exoplanet signal. The main
new feature of this instrument compared to other interferometric
instrument is the simultaneous use of modal filters (optical fibers)
and a dispersed fringe combiner using a spatial coding. The AMBER team
has therefore investigated carefully a data processing strategy for this
instrument and is providing a new type of data reduction method.   

Given the astonishingly quick evolution of ground based optical
interferometers in only two decades, in terms of baseline lengths and
number of recombined telescopes, the interest of using the practical
characteristics of single mode fibers to carry and recombine the
light, as first proposed by \citet{connes_1} with his conceptual FLOAT
interferometer, is now well established. Furthermore, in the light of the
FLUOR experiment on the IOTA interferometer, which demonstrated for
the first time the ``on-sky'' feasibility of such interferometers,
\citet{foresto_1} showed that making use of single mode waveguides
could also increase the performances of optical interferometry, thanks
to their remarkable properties of spatial filtering which change the
phase fluctuations of the atmospheric turbulent wavefront into
intensity fluctuations. Indeed, by monitoring in real time these
fluctuations thanks to dedicated photometric outputs and by performing
instantaneous photometric calibration, he experimentally proved that
single mode interferometry could achieve visibility measurements with
precisions of $1\%$ or below. Achievement of such level of
performances has since been confirmed with the IONIC integrated optic
beam combiner set up on the same interferometer \citep{lebouquin_1}.

Surprisingly, the effect of single mode waveguides on the
interferometric signal has been only recently studied from a
theoretical point of view.  \citet{ruilier_1} showed through numerical
simulations in presence of partial correction by Adaptive Optics that
spatial filtering provided a gain on the visibility signal to noise
ratio. However his study was limited to the case of a point source.
The case of sources with given spatial extent was first theoretically
addressed by \citet{dyer_christensen_1} from a geometrical point of
view. They proved that the visibility obtained from single mode
interferometry was biased, the object being multiplied by the antenna
lobe (the point spread function of one single telescope) exactly as it
happens in radio interferometry \citep{guilloteau_1}. An equivalent
geometrical bias was also characterized for the closure phase
\citep{longueteau_1}. Then \citet{guyon_1} noticed on his simulations
that took into account the presence of atmospheric turbulence, that
interferometric observations of extended objects (resolved by one
single telescope) could not be completely corrected from atmospheric
perturbations, therefore lowering the performances of single mode
interferometry. Finally, by thoroughly describing the propagation of
the electric field through single mode waveguides in the general case
of partial correction by Adaptive Optics and for a source with given
spatial extent, \citet{mege_1} unified previous studies and introduced
the concept of modal visibility, which in the general case does not
equal the source visibility $V_o$ and exhibits a jointly geometrical
and atmospheric bias. Nevertheless they also showed that for compact
sources, i.e. smaller than one Airy disk, the mutual coherence factor
$\mu$ could be written under the form of a simple product $\mu =
T_iT_aV_o$ where $T_i$ and $T_a$ are respectively the instrumental and
the atmospheric transfer functions which can be calibrated.
Recently, \citet{tatulli_1} deduced from an analytical approach
that in the specific case of compact objects, the benefit brought by
single mode waveguides is substantial, not only in terms of signal to
noise ratio of the visibility but also on the robustness of the
estimator.

Hence, following the path opened by the FLUOR experiment, the AMBER
instrument -- the three-beam combiner of the VLTI \citep{petrov_1} 
-- makes use of the filtering properties of single mode
fibers. However, on the contrary of FLUOR, PTI \citep{colavita_1}
or VINCI on the VLTI \citep{kervella_1} where the fringes are coded
temporally with a movable piezzo-electric mirror, the interference
pattern is scanned spatially thanks to separated output pupils which
separation fixes the spatial coding frequency of the fringes, as in
the case of the GI2T interferometer \citep{mourard_1}.  Thus, if data
reduction methods have already been proposed for single mode
interferometers using temporal coding \citep{colavita_2, kervella_2},
this paper is the first that presents a signal processing algorithm
dedicated to single-mode interferometry with spatial beam
recombination. Moreover, in the case of AMBER, the configuration of
the output pupils, that is the spatial coding frequency, imposes in
the three telescopes case a partial overlap of the
interferometric peaks in the Fourier plane.  As a consequence, the
data reduction based on the classical estimators in the Fourier plane
\citep{roddier_lena_1,mourard_2} cannot be performed.  The AMBER data
reduction procedure is based on an direct analysis in the detector
plane, which principle is an optimization of the "ABCD" estimator as
derived in \citet{colavita_2}. The specificity of the AMBER coding and its subsequent estimation of the observables arises from the will to
characterize and to make use of the linear relationship between the
pixels (i.e. the interferograms on the detector) and the observables
(i.e. the complex visibilities). In other words, the AMBER data
reduction algorithm is based on the modelling of the interferogram in
the detector plane.

In Sect.~\ref{sec_presentation}, we present the AMBER experiment
from a signal processing point of view and we introduce the
interferometric equation governing this instrument. We develop in
Sect.~\ref{sec_principles} the specific data reduction processes of
AMBER, and then derive the estimators of the interferometric
observables.
Successive steps of the data reduction method are given in Section
\ref{sec_soft}, as performed by the software provided to the
community. Finally, the data reduction algorithm is validated in
Sect.~\ref{sec_valid} through several ``on-sky'' observations with the VLTI (commissioning and Science Demonstration Time (SDT)). 
Present and future performances of this instrument are discussed.
  
\begin{figure*}[!t]
\begin{center}
\begin{tabular}{cc}
\includegraphics[height=5cm]{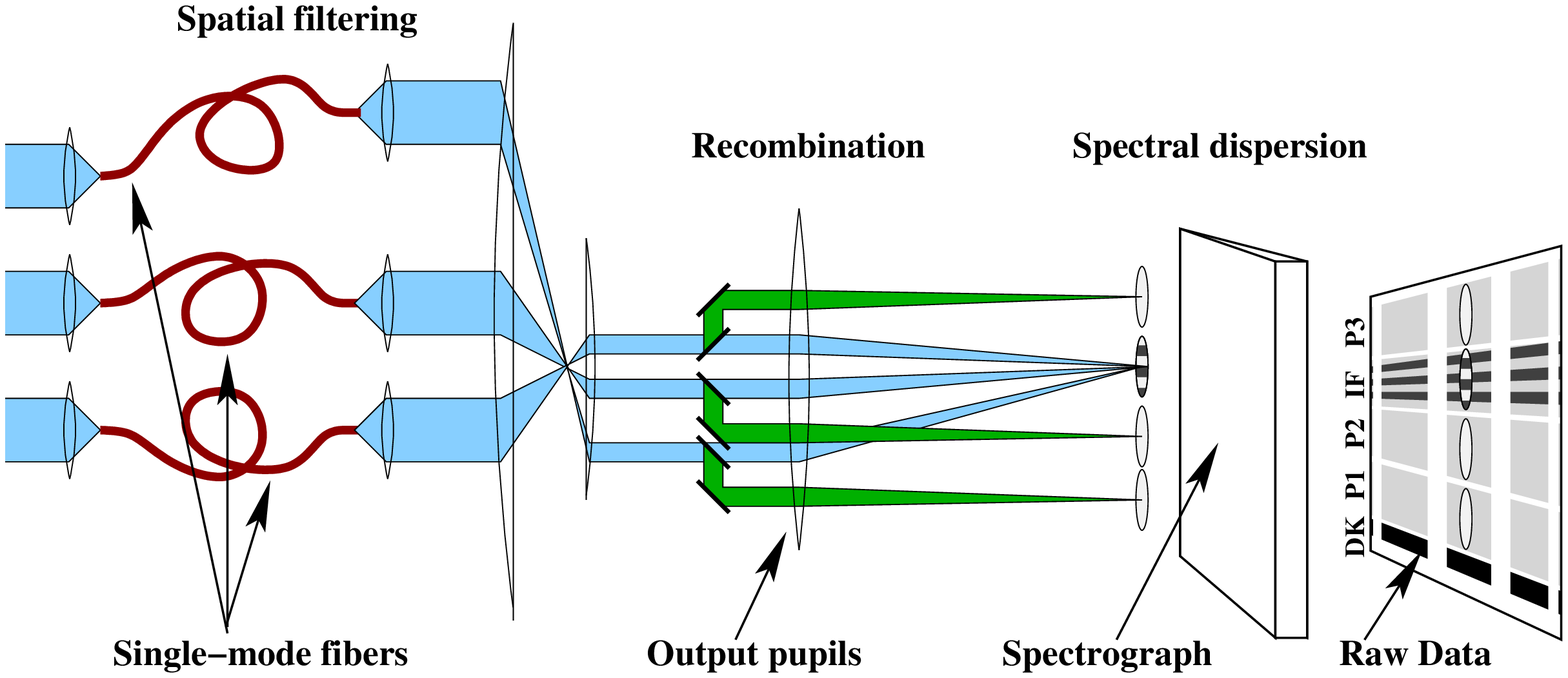} & 
\includegraphics[height=5cm]{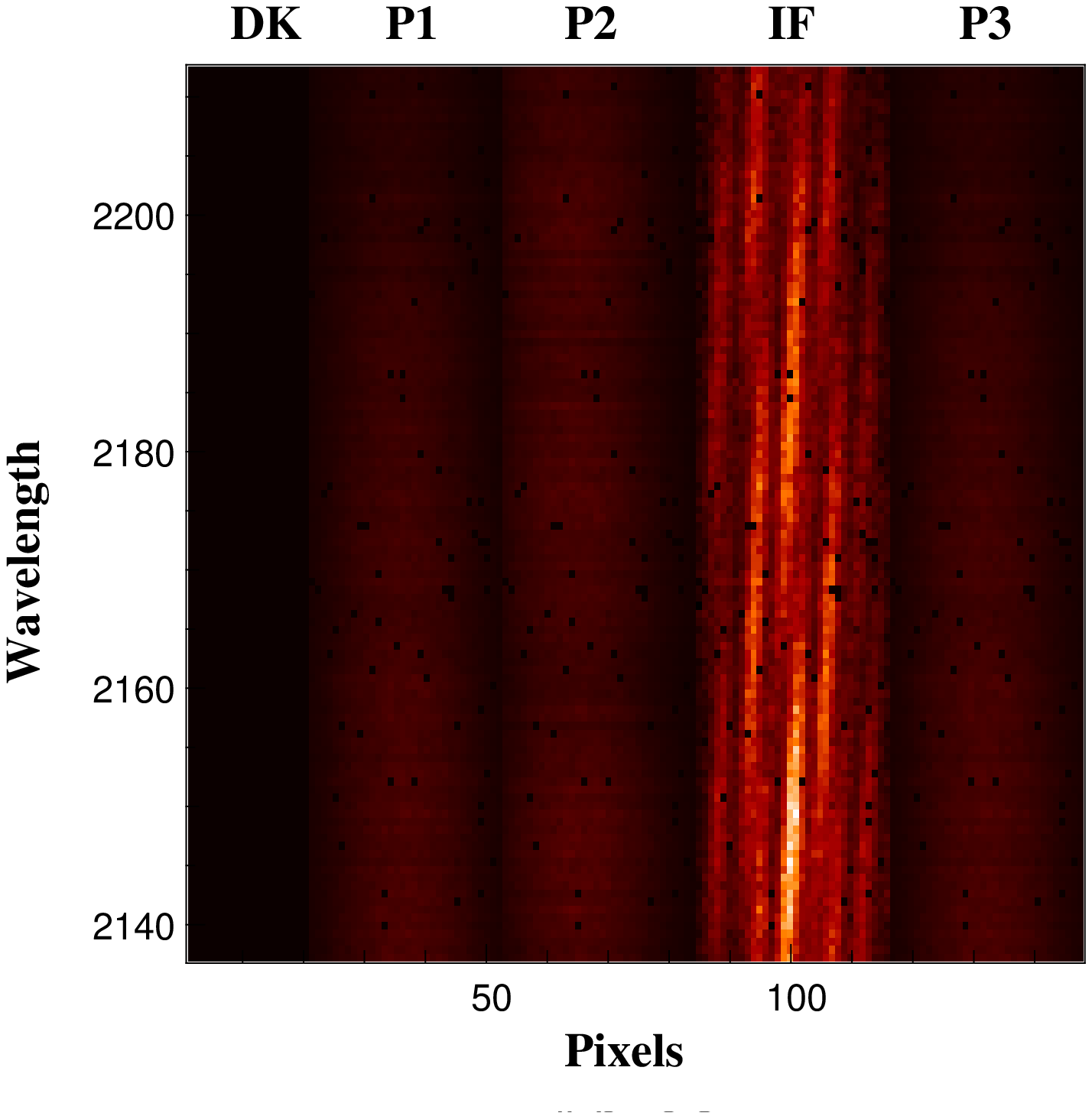} 
\end{tabular}
\caption{Left panel: Sketch of the AMBER instrument. The light enters the
  instrument from the left and is propagating from left to right, until the raw data are recorded on the detector. Further details are given in the text. 
	Right panel: AMBER reconstituted image from the raw data
  recorded during the 3-telescope observation of the calibrator \object{HD135382} in February 2005, in the medium spectral resolution mode. \texttt{DK} corresponds to a dark region,
  \texttt{Pk} are the vertically dispersed spectra obtained from each
  telescope and \texttt{IF} is the spectrally dispersed interferogram.
  \label{fig_amber}}
\end{center}
\end{figure*}
\section{Presentation of the instrument} \label{sec_presentation}
\subsection{Image formation} 
\label{sec_image}

The process of image formation of AMBER is sketched on
Fig.~\ref{fig_amber} (left) from a signal processing point of view. It
consists in three major steps. First, the beams from the three
telescopes are filtered by single mode fibers to convert phase
fluctuations of the corrugated wavefronts into intensity fluctuations
that are monitored. The fraction of light entering the fiber is
called the coupling coefficient \citep{shaklan_1} and depends on the
Strehl ratio \citep{foresto_2}.  At this point, a pair of conjugated
cylindrical mirrors compresses by a factor of about 12 the individual
beams exiting from fibers into one dimensional elongated beams to be
injected in the entrance slit of the spectrograph.  For each of the
three beams, beam-splitters placed inside the spectrograph select part
of the light and induce three different tilt angles so that each beam
is imaged at different locations of the detector. These are called
photometric channels and are each one relative to a corresponding
incoming beam.  The remaining parts of the light of the three beams
are overlapped on the detector image plane to form fringes. The spatial
coding frequencies of the fringes $f$ are fixed by the separation of
the individual output pupils.  They are respectively $f =
[1,2,3]d/\lambda$, where $d$ is the output pupil diameter.  Since the
beams hit a spectral dispersing element (a prism glued on a mirror or
one of the two gratings) in the pupil plane, the interferogram
and the photometries are spectrally dispersed perpendicularly to the
spatial coding. The dispersed interferogram arising from the
beam combination, as well as the photometric outputs are recorded on the
infrared detector, which characteristics are given in
Table~\ref{table_det}.
\begin{table}[!t]
\begin{center}
\caption{\label{table_det} Detector properties.}
\begin{tabular}{cc}
\hline \hline
\multicolumn{2}{c}{Detector specifications} \\
\hline \hline
Society/Name & Rockwell/HAWAII\\
Composition & HgCdTe \\
Number of pixels & $512 \times 512$\\
Pixel size & $18.5\mu\mathrm{m} \times 18.5\mu\mathrm{m}$\\
Spectral width & $0.8\mu\mathrm{m} - 2.5\mu\mathrm{m}$\\
Readout  noise &  $9\mathrm{e}^{-}$\\
$e^-/ADU$ & 4.18\\
Cooling  & Liquid nitrogen\\
Temperature & $78\mathrm{K}$\\
Autonomy of cryostat& $24\mathrm{h}$\\\hline
\end{tabular}
\end{center}
\end{table}

The detector consists in a 512 x 512 pixel array with the vertical dimension
aligned with the wavelength direction. The first 20 pixels of each scanline
of the detector are masked and never receive any light, allowing  to
estimate the readout noise and bias during an exposure.
The light from the two (resp. 3) telescopes comes in three (resp. 4) beams,
one "interferometric" beam where the interference fringes are located, and
two (resp. three) "photometric" beams. These 3 (resp. 4)  beams are
dispersed and spread over three (resp. 4) vertical areas on the detector.
The detector is read in subwindows. Horizontally, these subwindows are
centered on the regions where the beams are dispersed, with a typical width
of 32 to 40 pixels. Vertically, the detector can be set up to read up to
three subwindows (covering up to three different wavelength ranges).  The
Raw Data format used by AMBER records individually these subframes. However,
as sketched in the right panel of Fig.~\ref{fig_amber}, 
the AMBER Raw Data can be conceived as the grouping together of these subwindows:
\begin{itemize}
\item the left column (noted ``DK'' ) , contains the masked pixels.
\item the two following columns (noted ``P1'' and ``P2''), and the
 right one in the three telescope mode (noted ``P3''), of usually $32$
pixels wide,
 are the photometric outputs. They record the photometric signal coming from
 the three telescopes. When dealing with 2-telescope observations, only channels P1 and P2 are enlighted.
\item the fourth column is the interferometric output (noted ``IF''). it
exhibits the interference fringes arising from the recombination of the beams  (that is two or three beams, according to the number of telescopes used). We call $N_{pix}$ the number of pixels in this column, which
is usually $N_{pix}=32$.
\end{itemize}
The individual image which is recorded during the detector
integration time (DIT) is called a {\it frame}. A cube of frames
obtained during the exposure time is called an {\it exposure}.

\subsection{AMBER interferometric equation}

The following demonstration is given considering a generic $N_{tel}
\ge 2$ telescope interferometer. In the specific case of AMBER
however, $N_{tel} =2$ or $N_{tel} =3$ only. Each line of the detector
being independent with respect to each other, we can focus our
attention on one single spectral channel\footnote{In practice, there is a  previous image centering step, where each channel is re-centered with respect to each other along the wavelength dimension, as explained in Sect.~\ref{sec_soft}}
, which is assumed here to be
monochromatic. The effect of a spectral bandwidth on the
interferometric equation is treated in Section
\ref{sec_spectralcoherence}.

{\bf Interferometric output:} when only the $i^{th}$ beam is
illuminated, the signal recorded in the interferometric channel is the
photometric flux $F^i$ spread on the Airy pattern $a^i_k$, that is the
diffraction pattern of the $i^{th}$ output pupil weighted by the
single mode of the fiber. $k$ is the pixel number on the detector, $\alpha$ being the associated angular variable.
$F^i$ results in the total source photon flux $N$ attenuated by the
total transmission of the $i^{th}$ optical train $t^{i}$, i.e. the
product of the optical throughput (including atmosphere and optical
train of the VLTI and the instrument) and the coupling coefficient of
the single mode fiber:
\begin{equation}
F^i = N t^{i} \label{eq_fi}
\end{equation}

When  beams $i$ and $j$ are illuminated simultaneously, the coherent
addition of both beams results in an interferometric component
superimposed to the photometric continuum. The interferometric part,
that is the fringes, arises from the amplitude modulation of the coherent flux
$F_c^{ij}$ at the coding frequency $f^{ij}$. The coherent flux is the
  geometrical product of the photometric fluxes, weighted by the visibility:
\begin{equation}
F_c^{ij} = 2 N \sqrt{t^it^j} V^{ij} \mathrm{e}^{i(\Phi^{ij} +
  \phi_p^{ij})} \label{eq_fc}
\end{equation} 
where  $V^{ij}\mathrm{e}^{i\Phi^{ij}}$ is the complex modal 
visibility \citep{mege_1} and $\phi_p^{ij}$ takes into account a
potential differential atmospheric piston. Note that strictly
speaking the modal visibility is not the source visibility. However
the study of the relation between the modal visibility and the source
visibility is beyond the scope of this paper, and further informations
can be found in \citet{mege_1} and \citet{tatulli_1}. 
Here we consider our observable to be the complex modal visibility.

Such an analysis can be done for each pair of beams arising from  the
interferometer. 
As a result, the interferogram recorded on the detector can be written
in the general form:
\begin{equation}
i_k = \sum_{i}^{N_{tel}} a^i_k F^i +  \sum_{i <j}^{N_{tel}} \sqrt{a^i_ka^j_k}C^{ij}_{B}\mathrm{Re}\left[F_c^{ij}\mathrm{e}^{i(2\pi\alpha_k{f^{ij}} + \phi^{ij}_s + \Phi^{ij}_B)}\right]\label{eq_interf_general}
\end{equation}
$\phi^{ij}_s$ is the instrumental phase taking
into account possible misalignment and/or differential phase
between the beams $a^i_k$ and $a^j_k$. $C^{ij}_{B}$ and $\Phi^{ij}_B$ are respectively the loss of contrast and the phase shift due to polarization mismatch between the two beams (after the polarizers), as rotation of the single mode fibers might induce. 
This equation is governing the AMBER fringe pattern, that is the interferometric channel of the fourth column.
The first sum in Eq. (\ref{eq_interf_general}), which represents the continuum part of the interference pattern, is called from now on the DC component, and the second sum, which describes the high frequency part (that is the coded fringes), is called the AC component of the interferometric output. \\ \\ 
{\bf Photometric  outputs:} 
thanks to the photometric channels, the number of photoevents
$p^i(\alpha)$ coming from each telescope can be estimated independently:
\begin{equation}
p^i_k =  F^i b^i_k \label{eq_photo_general}
\end{equation}
where $b^i_k$ is the beam profile in the $i^{th}$ photometric channel. The previous equation rules the photometric channels.

\section{Data reduction algorithm} \label{sec_principles}
The AMBER data reduction algorithm
is based on the modelling of the interferogram in the detector
plane. Such a method requires an accurate calibration of the instrument.
 
\subsection{Modelling the interferogram}
In order to model the interferogram, we discriminate between
the astrophysical and instrumental parts in the interferometric
equation. It comes
\begin{equation}
i_k = \sum_{i}^{N_{tel}}  a^i_{k} F^i + \sum_{i < j}^{N_{tel}}
\left[c_k^{ij}{R^{ij}}+d_k^{ij}{I^{ij}}\right]  \label{eq_ik}
\end{equation}
with
\begin{equation}
c_k^{ij}=C^{ij}_{B}\frac{\sqrt{a^i_{k}a^j_{k}}}{\sqrt{\sum_k a_k^{i} a_k^{j}}}\cos(2\pi\alpha_k{f^{ij}}+ \phi^{ij}_s + \Phi^{ij}_B) \label{eq_ck}
\end{equation}
\begin{equation}
d_k^{ij}=C^{ij}_{B}\frac{\sqrt{a^i_{k}a^j_{k}}}{\sqrt{\sum_k a_k^{i} a_k^{j}}}\sin(2\pi\alpha_k{f^{ij}}+ \phi^{ij}_s + \Phi^{ij}_B) \label{eq_dk}
\end{equation}
and 
\begin{equation}
{R^{ij}} = \sqrt{\sum_k a_k^{i} a_k^{j}} \mathrm{Re}\left[F_c^{ij}\right],~~{I^{ij}} = \sqrt{\sum_k a_k^{i} a_k^{j}} \mathrm{Im}\left[F_c^{ij}\right] \label{eq_Rij}
\end{equation}
As an analogy with telecom data processing, $c_k^{ij}$ and $d_k^{ij}$ are called the \emph{carrying waves} of the signal
at the coding frequency $f^{ij}$, since they \emph{carry} (in terms of
amplitude modulation) 
$R^{ij}$ and $I^{ij}$, which are directly linked to the complex coherent
flux (as shown by Eq. (\ref{eq_Rij})).

The estimated photometric fluxes $P^i$ are computed from the photometric
channels (see Eq. (\ref{eq_photo_general})): 
\begin{equation}
P^i = F^i \sum_k b_k^i \label{eq_Pi}
\end{equation}
If we know the ratio $v_k^i$ --which depends only of the instrumental configuration -- between the measured photometric fluxes $P^i$
and the corresponding DC components of the interferogram, we can have an estimation
of the latter thanks to the following formula:
\begin{equation}
 a^i_{k} F^i = P^i v_k^i \label{eq_estimFi}
\end{equation}
We then can compute the DC continuum corrected interferogram $m_k$:
\begin{equation}
m_k = i_k - \sum_{i=1}^{N_{tel}} P^i v_k^i \label{eq_calculmk}
\end{equation}
which can be rewritten:
\begin{equation}
m_k = c_k^{ij}R^{ij} - d_k^{ij}I^{ij}
\end{equation} 
This equation defines a system of $N_{pix}$ linear equations with
$2N_b=N_{tel}(N_{tel}-1)$ unknowns (i.e. twice the number of baselines). 
It characterizes the linear link
between the pixels on the detector and the complex visibility:
\begin{equation} 
\left(\begin{array}{c} m_1 \\ \vert \\ m_{N_{pix}} \end{array}\right)  = \overbrace{\left(\begin{array}{ccc}
 \unldots&  c_1^{ij} &\unldots\\ \vert &\deuxvdots&\vert\\
  \unldots  &c_{N_{pix}}^{ij} &\unldots
   \end{array}\right.}^{N_b}\overbrace{\left. \begin{array}{ccc}
 \unldots&  d_1^{ij} &\unldots\\ \vert &\deuxvdots & \vert\\
  \unldots  &d_{N_{pix}}^{ij} &\unldots
   \end{array}\right)}^{N_b}
 \left(\begin{array}{c} \deuxvdots
    \\ R^{ij} \\ \deuxvdots  \\ I^{ij} \\ \deuxvdots \end{array} \right) = \mathrm{V2PM} \left(\begin{array}{c} \deuxvdots
    \\ R^{ij} \\ \deuxvdots  \\ I^{ij} \\ \deuxvdots \end{array} \right) \label{eq_fitfringe}
\end{equation}  
The $\mathrm{V2PM}$ matrix (namely {\it Visibility To Pixel Matrix}), which contains the carrying waves, holds the information
about the interferometric beams $\sqrt{a^i_{k}a^j_{k}}$, the coding
frequencies $f^{ij}$ and the instrumental differential phases
$\phi^{ij}_s$. Together with the $v_k^i$, they entirely describe the
instrument from a signal processing point of view. These quantities,
namely $c_k^{ij}$, $d_k^{ij}$ and $v_k^i$ have however to be
calibrated.


\subsection{Calibration procedure} \label{sec_calib}
\begin{table}[!t]
\begin{center}
\caption{\label{tab_calib}Acquisition sequence of calibration files}
\begin{tabular}{cccccc}
\hline \hline
Step & Sh $1$ & Sh $2$ & Sh $3$ & Phase $\gamma_0$ & DPR key \\
\hline \hline
1 &  O & X & X & NO & 2P2V, 3P2V\\
2 &  X & O  & X & NO & 2P2V, 3P2V\\
3 &  O & O & X & NO & 2P2V, 3P2V\\
4 &  O & O & X & YES & 2P2V, 3P2V\\
\hline
5 &  X & X & O  & NO & 3P2V\\
6 &  O & X & O & NO & 3P2V\\
7 &  O & X & O & YES & 3P2V\\
8 &  X & O & O & NO & 3P2V\\
9 &  X & O & O & YES & 3P2V\\
\hline \hline
\end{tabular}\\
Sh = Shutter; O = Open; X = Closed
\end{center}
\end{table}
\begin{figure*}[!t]
\begin{center}
\begin{tabular}{ccc} 
\includegraphics[height=5cm]{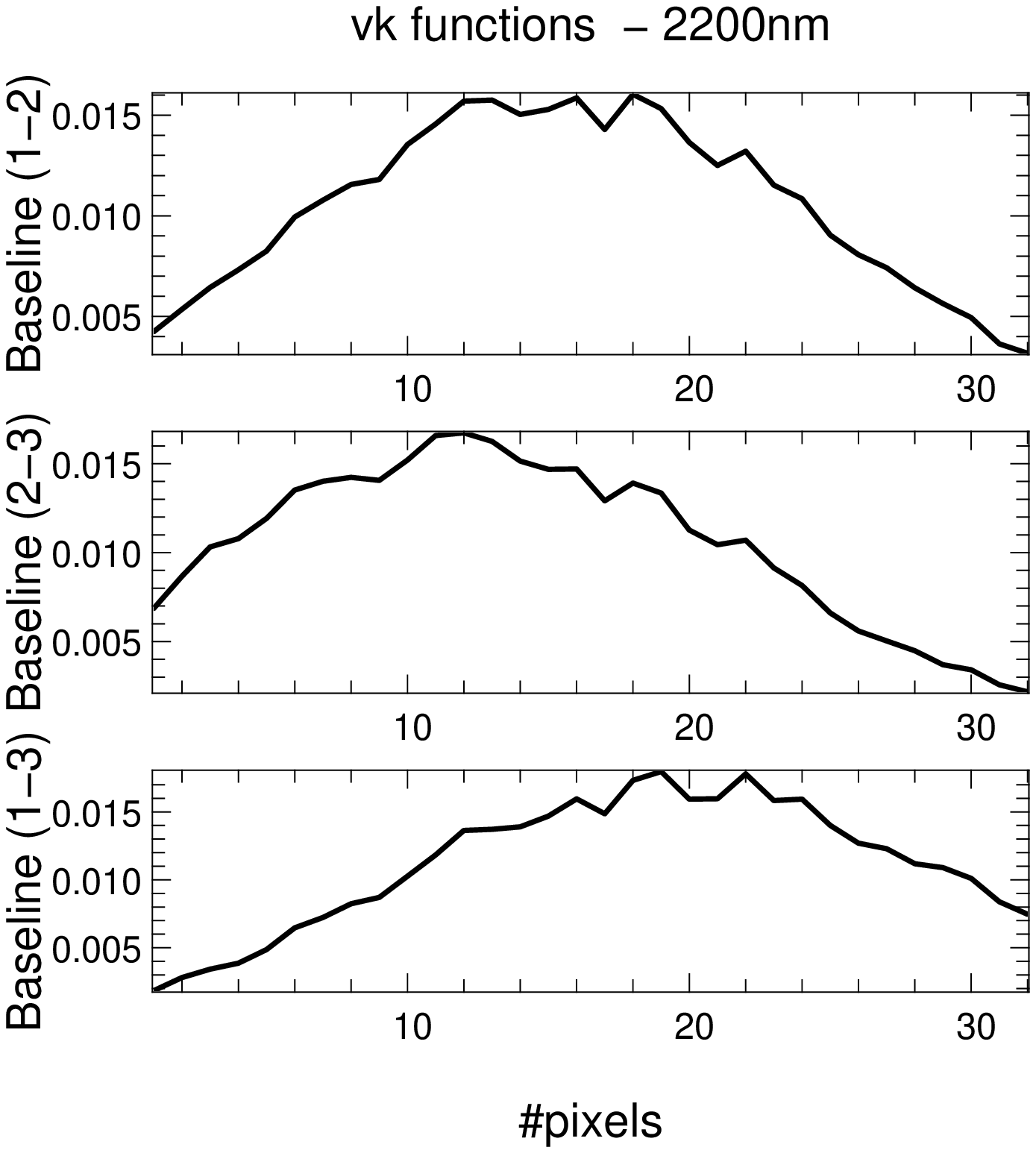} &
\includegraphics[height=5cm]{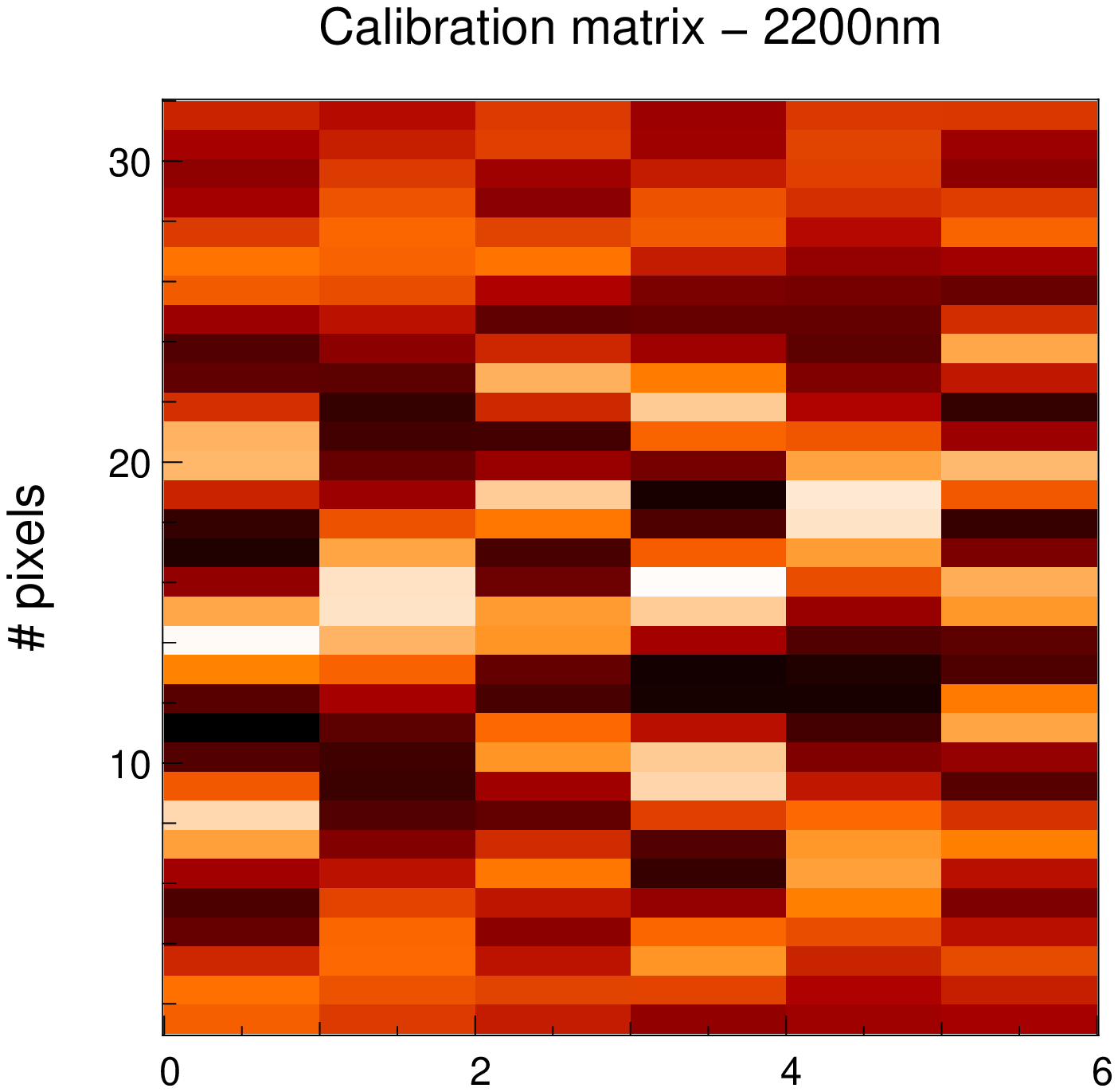} & 
\includegraphics[height=5cm]{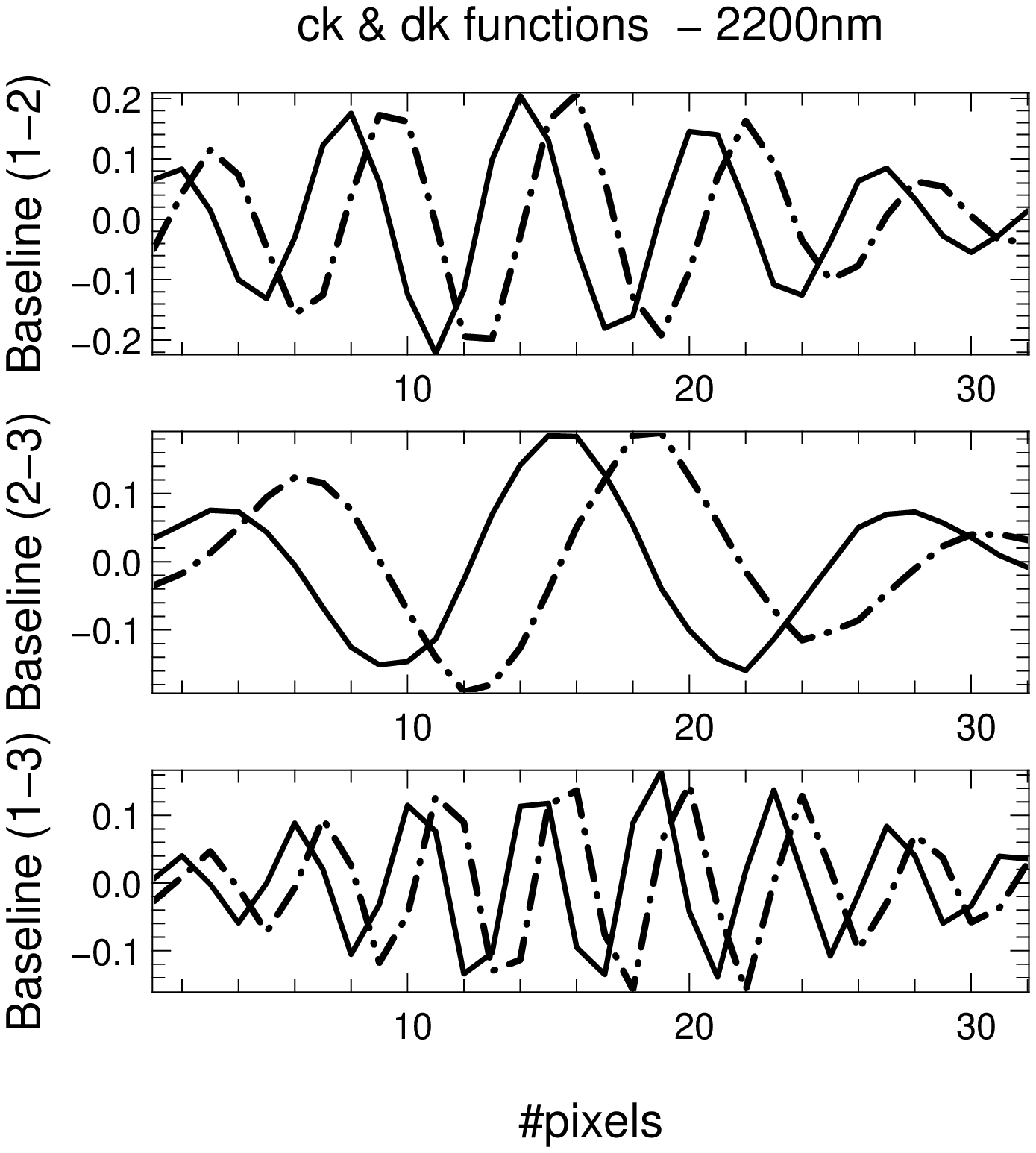} 
\end{tabular}
\caption{\label{fig_p2vm} Outputs of the calibration procedures. Examples have been chosen for one given wavelength: $\lambda = 2.2\mu\mathrm{m}$. Left: the $v_k^i$ functions. Middle: the matrix containing the carrying waves, the first three columns are the $c_k^{ij}$ functions for each baseline, and the three last columns are the respective $d_k^{ij}$ functions. One can see that for each baseline $c_k^{ij}$ and $d_k^{ij}$ are in quadrature. Right: another representation of the carrying waves. From top to bottom, both sinusoidal functions correspond respectively to columns $1-4$, $2-5$ and $3-6$ of the calibration matrix.}
\end{center}
\end{figure*}

The calibration procedure is performed thanks to an internal source
located in the \emph{Calibration and Alignment Unit} (CAU) of AMBER
\citep{petrov_1}. It consists in acquiring a sequence of high
signal-to-noise ratio calibration files, which successive
configurations are summarized in Table \ref{tab_calib} and detailed
below. Since the calibration is done in laboratory, the desired level
of accuracy of the measurements is insured by choosing the appropriate
integration time.  As an example, typical integrations time in ``average accuracy''  mode are (for the full calibration process) $\tau = 17\mathrm{s},~30\mathrm{s},~800\mathrm{s}$ for respectively low, medium and high spectral resolution modes in the K band, and $100$ times higher for the ``high accuracy'' calibration mode.

The sequence of calibration files has been chosen to accommodate both
two and three-telescope operations. For a two-telescope operation, only the 4
first steps are needed. Raw data FITS files produced by the ESO
instruments bear no identifiable name and can only be identified as,
e.g., files relevant to the calibration of the V2PM matrix, 
only by the
presence of dedicated FITS keywords (ESO's pipeline Data~PRoduct keys
or ``DPR keys'') in their header. The DPR keys used are listed in
Table~\ref{tab_calib}.

First (steps $1$ and $2$---and $5$ when in 3-telescope mode), for each
telescope beam, an image is recorded with
only this shutter opened. The fraction of flux
measured between the interferometric channel and the illuminated
photometric channel leads to an accurate estimation of the $v_k^i$
functions.  Then, in order to compute the carrying waves $c_k^{ij}$
and $d_k^{ij}$, one needs to have two independent (in terms of
algebra) measurements of the interferogram since there are two
unknowns (per baseline) to compute. The principle is the following:
two shutters are opened simultaneously (respectively steps $3/4$, $6/7$,
and $8/9$) and for each pair of beams, the interferogram is recorded on
the detector. Such an interferogram corrected from its DC component
and calibrated by the photometry yields  the knowledge of the
$c_k^{ij}$ carrying wave. To obtain its quadratic counterpart, the
previous procedure is repeated by introducing a known phase shift close to $90$ degree 
$\gamma_0$ using piezoelectric mirrors at the entrance of beams 2 and 3. Computing the
$d_k^{ij}$ function from the knowledge of $c_k^{ij}$ and
$\gamma_0$ is straightforward. 
Note that by construction: (i) the carrying waves are computed with the unknown  system phase $\Phi^c$ (possible phase of the internal source, differential optical path difference introduced at the CAU level, etc...), and that; (ii) 
since the internal source in the CAU is slightly 
resolved by the largest baseline (1--3) of the output pupils, the
carrying waves for this specific baseline are weighted by the
visibility $V_c$ of the internal source. Hence at this point, the carrying waves are following expressions slightly different from their original definition given by Eq.'s (\ref{eq_ck}) and (\ref{eq_dk}):
\begin{equation}
c_k^{ij}=\frac{\sqrt{a^i_{k}a^j_{k}}}{\sqrt{\sum_k a_k^{i} a_k^{j}}}C^{ij}_{B}V_c^{ij}\cos(2\pi\alpha_k{f^{ij}}+ \phi^{ij}_s +  \Phi^{ij}_B + \Phi^{ij}_c) 
\end{equation}
\begin{equation}
d_k^{ij}= \frac{\sqrt{a^i_{k}a^j_{k}}}{\sqrt{\sum_k a_k^{i} a_k^{j}}}C^{ij}_{B}V_c^{ij}\sin(2\pi\alpha_k{f^{ij}}+ \phi^{ij}_s +  \Phi^{ij}_B + \Phi^{ij}_c) 
\end{equation}
However, since $c_k^{ij}$ and $d_k^{ij}$ are shifted
by $\pi/2$, they insure the following relation:
\begin{equation}
\sum_k^{N_{pix}} c_k^2 + d_k^2 = C_{B}^2V_c^2 \label{eq_ck2_dk2}
\end{equation}
Hence the conjugated loss of visibility due to the internal source and 
the polarization effects can be known and calibrated\footnote[1]{This step is not yet provided in the amdlib software described in Sect.~\ref{sec_soft}} by computing previous formula. Unfortunately, since it is not possible to disentangle between both contrast losses, and since the $V_c$ factor only affects the interferograms arising from the calibration procedure, and not from the observation, the visibility estimated on a star will be affected from this factor as well, as shown in Section \ref{eq_squaredvis}.  

\begin{figure}[!t]
\begin{center}
\begin{tabular}{ccc}
\includegraphics[height=5cm]{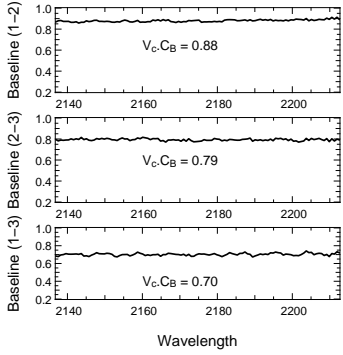} &
\end{tabular}
\caption{\label{fig_sumckdk}Contrast loss due to polarization effects and partial resolution of the internal source, as a function of the wavelength. The 3-telescope P2VM used is the same than the one presented in Fig. \ref{fig_p2vm}. The errors bars are roughly at the level of the contrast loss rms along the wavelength. In other words, the  contrast loss is constant over the wavelength range.}
\end{center}
\end{figure}

Figure \ref{fig_sumckdk} illustrates Eq. \ref{eq_ck2_dk2}. For the baselines $(1,2)$ and $(2,3)$, the contrast loss arises from polarization effects since the internal source is unresolved. We find respectively $C_{B}^{12} \simeq 0.9 $, and $C_{B}^{23}\simeq 0.8$. For the third baseline $(1,3)$, the internal source is partially resolved, which explain an higher contrast loss, $C_{B}^{13}V_c^{13}\simeq 0.7$. 

\subsection{Fringe fitting}
To estimate the coherent fluxes $R^{ij}$ and $I^{ij}$ which are at the basis of the computation of the whole AMBER observables, one has to solve the inverse problem described by Eq. (\ref{eq_fitfringe}), that is one has to perform a $\chi^2$ linear fit of the fringes, with the coherent fluxes being the unknown parameters. The solution is given by the following equation:
\begin{equation}
[\widetilde{R}^{ij}, \widetilde{I}^{ij}] = \mathrm{P2VM} [m_k] \label{eq_fringefit}
\end{equation} 
where 
\begin{equation}
\mathrm{P2VM} = [\mathrm{V2PM}^T\mathcal{C}_M^{-1}{\bf \mathrm{V2PM}}]^{-1}\mathrm{V2PM}^T\mathcal{C}_M^{-1} \label{eq_invmat}
\end{equation} 
is the generalized inverse of the $\mathrm{V2PM}$ matrix, $\mathcal{C}_M$ being the covariance matrix of the measurements $m_k$, and $X^T$ denoting the transpose of the $X$ matrix.  P2VM means {\it Pixel to Visibility Matrix} 
since it allows to estimate the
complex visibility from the interferogram recorded on the detector.
Assuming that the pixels on the detector are uncorrelated, the $\mathcal{C}_M$ matrix is diagonal, with each term of the diagonal being defined by the variance of the DC corrected interferogram $\sigma^2(m_k)$. The fundamental error on the DC corrected interferogram 
arises from the photon noise and detector noise (of variance $\sigma$) 
corrupting the measurements, that is each pixel of the interferogram $i_k$, 
and the estimated photometric fluxes $P^i$. It comes:
\begin{equation}
\sigma^2(m_k) = \overline{i_k}+\sigma^2 + \sum_{i=1}^{N_{tel}}
\left[\overline{P_{i}} + N_{pix}\sigma^2\right](v^{i}_{k})^2 \label{eq_sigma2mk}
\end{equation}  

\subsection{Fringe detection} \label{sec_fringedet}
Positive detection of fringes in the measurements requires at the same time enough flux entering the fibers and high enough fringe contrast, so that the fringes rise of the noise level. As a result, the computation of the signal to noise ratio of the coherent flux, which takes into account both of the parameters,  appears naturally as the relevant criterion to use. It writes:
\begin{equation}
\mathrm{SNR}^2(t) =  \frac{1}{N_b}\frac{1}{N_l}\sum_{b}^{N_b} \sum_{l}^{N_l}
\left[\left(\frac{{R^b}^2(l,t)}{\sigma^2_{R^b}}-1\right) + \left(\frac{{I^b}^2(l,t)}{\sigma^2_{I^b}}-1\right)\right] \label{eq_fringeSNR}
\end{equation}
$b$ being for sake of simplicity the baseline number which describes each couple of telescopes $(i,j)$, $N_b$ being the number of baselines, and $N_l$ being the number of spectral channels. 
In absence of fringes, the quantities ${R^b}^2$ and ${I^b}^2$ are tending toward $\sigma^2_{R^b}$ and $\sigma^2_{I^b}$ respectively, thus driving the fringe criterion toward $0$. At the opposite, the presence of fringes above the noise level, that is when ${R^b}^2(l,t) > \sigma^2_{R^b}$ and/or ${I^b}^2(l,t) > \sigma^2_{I^b}$ imposes the fringe criterion to be strictly superior to $0$ and is directly linked to the quality of the frames. It thus allows to operate a fringe selection prior to the proper estimation of the observables, which step can be useful for sets of data recorded in bad observational conditions, as shown in Section \ref{sec_fringecrit}.        

$\sigma^2_{R^b}$ and $\sigma^2_{I^b}$, that is the bias part of respectively $R^2$ and $I^2$, can be easily computed from the definition of the real and
imaginary part of the coherent fluxes, which are linear combinations
of the DC continuum corrected interferograms $m_k$. If
$\zeta^{b}_{k}$ and $\xi^{b}_{k}$ are the coefficient of the P2VM
matrix, $R^{b}$ and $I^{b}$ verify the respective following
equations:
\begin{equation}
R^{b} = \sum_{k=1}^{N_{pix}} \zeta^{b}_km_{k},~~ I^{b} =\sum_{k=1}^{N_{pix}} \xi^{b}_{k}m_{k} \label{eq_R12^i12}
\end{equation} 
It comes straightforward that:
\begin{equation}
\sigma^2_{R^b} = \sum_{k}(\zeta^{b}_k)^2\sigma^2(m_k);~ \sigma^2_{I^b} = \sum_{k} (\xi^{b}_k)^2\sigma^2(m_k)  \label{eq_biasR2I2}
\end{equation}

\subsection{Estimation of the observables}

For each spectral channel, squared visibility and closure phase (in
the three telescope case) can be estimated from the interferogram.
Taking advantage of the spectral dispersion, differential phase can be
computed as well. In the following paragraphs, we denote with $\left<...\right>$ the
ensemble average of the different quantities. This average can be
performed either on the frames within an exposure and/or on the wavelengths.

\subsubsection{The squared visibility} \label{eq_squaredvis}
Theoretically speaking, the squared visibility is given by computing the ratio between the squared coherent flux and the photometric fluxes. 
Following Eq.'s~(\ref{eq_fi}), (\ref{eq_fc}), (\ref{eq_Rij}) and (\ref{eq_Pi}) it comes:
\begin{equation}
\frac{|F_c^{ij}|^2}{4F^iF^j} =  \frac{{R^{ij}}^2 + {I^{ij}}^2}{4P^iP^j\sum_{k}v_k^iv_k^j} = \frac{|V^{ij}|^2}{{V_c^{ij}}^2}
\end{equation}
Note that, thanks to the calibration process, the computed visibility is free from the instrumental contrast, that is the loss of contrast due to the instrument, but as mentioned in Section \ref{sec_calib}, the object visibility is weighted by the visibility of the internal source. However this factor (${V_c^{ij}}^2$) automatically disappears when doing the necessary atmospheric calibration (see Section \ref{sec_jitter}).

As a result the visibility -- atmospheric issues apart -- has still to be calibrated by observing a reference source.  

In practice, because data are noisy, we perform an ensemble average on the frames that compose the data cube (see Section \ref{sec_image}) to estimate the expected values of the square coherent flux and  the photometric fluxes, respectively. Taking the average of the squared modulus of the
coherent flux, that is doing a quadratic estimation, allows to handle
the problem of the random differential piston $\phi_p^{ij}$, but
introduces a quadratic bias due to the zero-mean photon and detector
noises \citep{perrin_1}. 
The expression of the squared visibility estimator, unbiased by
fundamental noises is therefore:
\begin{equation}
\frac{\widetilde{|V^{ij}|^2}}{{V_c^{ij}}^2} = \frac{\left<{R^{ij}}^2 + {I^{ij}}^2\right> -~\mathrm{Bias}\{{R^{ij}}^2 + {I^{ij}}^2\}}{4\left<P^iP^j\right>\sum_{k}v_k^iv_k^j}  \label{eq_v2}
\end{equation}  
The quadratic bias of the squared amplitude of the coherent flux writes as the quadratic sum of the biases of $R^2$ and $I^2$. From Eq. \ref{eq_biasR2I2}, it comes:
\begin{equation}
\mathrm{Bias}\{{R^{ij}}^2 + {I^{ij}}^2\} = \sum_{k}\left[(\zeta^{ij}_k)^2+(\xi^{ij}_k)^2\right]\sigma^2(m_k) \label{eq_biasRI2}
\end{equation}
Previous equation is nothing but the mathematical expression that
describes the bias as the quadratic sum of the errors of the
measurements $\sigma^2(m_k)$ (as defined by Eq. (\ref{eq_sigma2mk})), 
projected on the real and imaginary axis of the coherent flux. 

Using the squared visibility estimator of Eq. (\ref{eq_v2}), the theoretical error bars on the squared visibility can be computed from its second order Taylor expansion \citep{papoulis_1, kervella_2}:
\begin{equation}
\sigma^2(|V^{ij}|^2) = \frac{1}{M}\left[\frac{\sigma^2(|C^{ij}|^2)}{\overline{|C_{
ij}|^2}^2} + \frac{\sigma^2(P^iP^j)}{\overline{P^iP^j}^2}\right]\overline{|V^{ij}|^2}^2
\end{equation}
where $|C^{ij}|^2 = {R^{ij}}^2 + {I^{ij}}^2 -~\mathrm{Bias}\{{R^{ij}}^2 + {I^{ij}}^2\}$ is the unbiased squared coherent flux.
In practice, the expected value and the variance of the squared coherent flux and the photometric fluxes are computed empirically from the $M$ available measurements. It comes the following semi-empirical formula:
\begin{eqnarray}
\sigma_{stat}^2(\widetilde{|V^{ij}|^2}) &=& \frac{1}{M} \left[\frac{\left<|C^{ij}|^4\right>_M-\left<|C^{ij}|^2\right>^2_M}{\left<|C^{ij}|^2\right>^2_M}  \right. \nonumber \\ 
&+& \left.\frac{\left<{P^i}^2{P^j}^2\right>_M-\left<P^iP^j\right>^2_M}{\left<P^iP^j\right>^2_M}\right]\widetilde{|V^{ij}|^2}^2 \label{eq_sigv2_semiemp}
\end{eqnarray}

Note finally that, although quadratic estimation of the visibility has been
computed, the squared visibility will be systematically decreased by
the atmosphere jitter during the frame integration time. We focus on this effect in Sect.~\ref{sec_biasvis}.

\subsubsection{The closure phase}

By definition, the closure phase is the phase of the so-called
bispectrum ${B}^{123}$. The latter results on the ensemble average of
the coherent flux triple product and then estimated as follows:
\begin{equation}
\widetilde{B}^{123} = \left<C^{12}C^{23}{C^{13}}^{\ast}\right>  \label{eq_bispectre}
\end{equation}
where $C^{ij} = R^{ij} + i I^{ij}$. The closure phase then comes straightforward:
\begin{equation}
\widetilde{\phi_B}^{123} =
\mathrm{atan}\left[\frac{\mathrm{Im}(\widetilde{B}^{123})}{\mathrm{Re}(\widetilde{B}^{123})}\right]
\label{eq_cloture}
\end{equation}
The closure phase presents the advantage to be independent of the
atmosphere (e.g. \citet{roddier_1}).
However in the case of AMBER, the closure phase of the
image might not coincide with the one of the object and might be
biased because of the calibration process. If the so-called system phase
presents an non zero closure phase
$\Phi_c^{12}+\Phi_c^{23}-\Phi_c^{13}$, this
bias must be calibrated by observing a point source or at least a
centro-symmetrical object. So far, no theoretical computation of the error of the closure phase has been provided for the AMBER data reduction algorithm. 
Thus, closure phases internal error bars (i.e. that does not include systematics errors) are computed statistically, that is 
by taking the root mean square of all the individual frames, then divided by the square root of the number of frames, as it is illustrated in Section \ref{sec_illusclosure}.
  
\subsubsection{The differential phase}

The differential phase is the phase of the so-called cross spectrum
${W}_{12}$. For each baseline, the latter is estimated from the
complex coherent flux taken at two different wavelengths $\lambda_1$
and $\lambda_2$:

\begin{equation}
\widetilde{W_{12}^{ij}} =  \left<C^{ij}_{\lambda_1}{C^{ij}_{\lambda_2}}^{\ast}\right> \label{eq_interspectre} \\
\end{equation}
And the differential phase is:
\begin{equation}
\widetilde{\Delta\phi_{12}^{ij}} =
\mathrm{atan}\left[\frac{\mathrm{Im}\left(\widetilde{W_{12}^{ij}}\right)}{\mathrm{Re}\left(\widetilde{W_{12}^{ij}}\right)}\right]
\label{eq_phasedif}
\end{equation}
\subsubsection{The piston}\label{sec_piston} 
The interferometric  phase induced by the achromatic piston
term takes the form:

 \begin{equation}
   \phi^{ij}_\lambda = \frac{2 \pi \delta^{ij}}{\lambda} =
   {2 \pi \delta^{ij} \sigma}
   \label{eq_piston}
\end{equation}
where $\delta^{ij}$ is the achromatic differential piston between telescope i and j, $\lambda$ is the wavelength and $\sigma$ is the wavenumber (i.e. $\sigma=1/\lambda$).\\
{\bf First order Taylor expansion:}
At first order, the estimated differential phase of
Eq.~(\ref{eq_phasedif}) is a linear function which takes the generic
form $\Delta\phi_{12} = \phi_1 + 2\pi\left(\sigma_2-\sigma_1\right)\delta$.
Its slope
$\delta$ depends of the sum of atmospheric piston $\delta_p$ which
varies frame by frame, and of the linear component of the object differential phase $\delta_o$. A good estimate of this slope in
presence of noise is the argument of the average cross spectrum along
the wavelengths:
 \begin{equation}
   \widetilde{\delta^{ij}_p} + \widetilde{\delta^{ij}_o} = \frac  {{\rm arg}\left<W_{\lambda_{2l},\lambda_{2l+1}}^{ij} \right>_l}{2 \pi \left<\sigma_{\lambda_{2l+1}}-\sigma_{\lambda_{2l}}\right>_l }
   \label{eq_estimpiston}
\end{equation}
The estimation of the piston is unbiased when the wave number varies
linearly with the spectral pixel index (linear grating dispersion law). This
can be true with an excellent approximation at Medium Spectral Resolution and
High Spectral Resolution in the AMBER case. However, for the Low
Spectral Resolution, biases as high as 5\% in the estimation of piston
can occur. \\
{\bf Fitting the complex phasor:}
The achromatic piston can also be estimated from a least square fit of the
complex coherent flux. If we define the complex phasor as:
\begin{equation}
  \Psi_\lambda = C_\lambda^{ij} \times
  e^{
    \frac{
      2i \pi \delta^{ij}
    }
    {
      \lambda
    }
  }
  \label{eq_estimPistonPhaseur}
\end{equation}
$ \delta^{ij}$ can be retrieved by minimizing the phase of such complex phasor, or equivalently by minimizing the tangent of the phase. 
The $\chi^2$ is then defined as:
\begin{equation}
  \chi^2 = \frac{ \displaystyle \sum_\lambda \frac{ \left( \frac {\mathrm{Im}
          \left( \Psi_\lambda\right)}{\mathrm{Re}\left( \Psi_\lambda\right)}
      \right)^2 } { \sigma^2_{R_\lambda} + \sigma^2_{I_\lambda} } }
 { \displaystyle \sum_\lambda \frac{ 1 } { \sigma^2_{R_\lambda} +
     \sigma^2_{I_\lambda} }
  }
  \label{eq_chi2estimPistonPhaseur}
\end{equation}
This $\chi^2$ is highly non-linear and simple techniques
as gradient fitting cannot be used here. On the contrary, non-linear 
fitting techniques 
such as genetic or simulated annealing algorithms \citep{kirkpatrick_1}
must be used instead.


Note that, in order to disentangle between the atmospheric piston $\delta_p$ and the linear component of the differential phase $\delta_o$, the fitting techniques described above can be performed only using  spectral channels corresponding to the continuum of the source (i.e. outside spectral features), that is where its differential phase of the object is assumed to be zero.  

\subsection{Biases of the visibility} 
\label{sec_biasvis}

\subsubsection{Loss of spectral coherence} 
\label{sec_spectralcoherence}

The above derivation of the interferometric equation assumes a
monochromatic spectral channel. In practice the spectral width of one
spectral channel is non zero and depends on the resolution
$\mathcal{R}$ of the spectrograph.  As a consequence the coherence
length $\mathcal{L}_c$ of the interferogram is finite and equals
$\mathcal{L}_c = \lambda_0\mathcal{R}$ where $\lambda_0$ is the
reference wavelength in the spectral channel. Assuming a
linear decomposition of the phase of the interferogram and neglecting higher orders, the
interferogram is attenuated by a factor $\rho_k$, which can be
written:
\begin{equation}
\rho_k = \left|\widehat{\mathcal{F}}\left(\pi\frac{\delta_k + \delta_p + \delta_o}{\mathcal{L}_c}\right)\right|
\end{equation}
where $\widehat{\mathcal{F}}$ is the Fourier transform of the spectral
filter function. $\delta_k$ is the spatial sampling of the
interferogram, that is the pixel coordinates expressed in optical path difference (OPD) units, 
$\delta_p$ and $\delta_o$ being respectively the
atmospheric 
piston and the slope of the object spectral
differential phase, as defined in Sect.~\ref{sec_piston}.  Note that
for a square filter, the attenuation coefficient takes the well known
form of the sinc function:
\begin{equation}
\rho_k = \left|\mathrm{sinc}\left(\pi\frac{\delta_k + \delta_p + \delta_o}{\mathcal{L}_c}\right)\right| \label{eq_losscoherence}
\end{equation}     
In the low resolution mode where $\mathcal{R}=35$, the attenuation
coefficient severely depends on the pixel position $\delta_k$ which is calibratable quantity. Nonetheless, the compensation of this effect requires an iterative process in two steps where (i) the estimation of $\delta_p + \delta_o$ is performed as described in Section \ref{sec_piston}, and (ii) the  $\rho_k$ attenuation correction is applied directly to the DC corrected interferograms $m_k$. The loop is then repeated until convergence. This algorithm, which not yet implemented in the software, will be described in greater details in a further paper. 

In the medium and high resolution (where respectively
$\mathcal{R}=1500$ and $\mathcal{R}=10000$) however, the OPD $\delta_k$ 
due the spatial sampling of AMBER can be neglected. Indeed
this approximation drives to a relative error of the coefficient below
$10^{-3}$ and $10^{-5}$ respectively, that is within the specified
error bars of the visibility. In such a case, the loss of spectral
coherence simply results in biasing frame to frame the visibility by a
factor $\rho(\delta_p+\delta_o)$. This bias can be corrected by
knowing the shape of the spectral filter and by estimating the piston
$\delta_p+\delta_o$ thanks to Eq. (\ref{eq_estimpiston}).

\subsubsection{Atmospheric jitter} \label{sec_jitter}
Although a quadratic estimation of the visibility has been performed
to avoid the differential piston to completely cancel out the fringes,
the high frequency variations of the latter during the integration
time -- so called high-pass jitter -- nevertheless blur the fringes.
As a result, the coherent flux, hence the visibility, is attenuated. In
average, the attenuation coefficient $\Gamma$ of the squared visibility
is given by \citet{colavita_2}:
\begin{equation}
\Gamma = \exp(-\sigma^2_{\phi^p_{hf}})
\end{equation} 
where $\sigma^2_{\phi^p_{hf}}$ is the variance of the high-pass jitter
$\phi^p_{hf}$.

For the time being, this atmospheric effect is compensated by
calibrating the source visibility with a reference source observed
shortly before and after the scientific target to insure similar
atmospheric conditions. We have also planned to provide in the near future 
a more accurate
calibration of this effect, based on the computation of the variance
of the so-called ``first difference phase jitter'', that is the
difference of the average piston taken between two successive
exposures, as proposed by \citet{colavita_2} for the PTI
interferometer and successfully applied by \citet{malbet_2}.
However, jitter analysis (as it is illustrated in Section \ref{sec_fringecrit}) 
cannot be tested and validated as long as the
extra-sources of vibrations due to VLTI instabilities (delay lines,
adaptive optics...), hardly calibratable, are clearly
identified and suppressed. Note as well that the use of the accurate
fringe tracker FINITO \citep{gai_1}, soon expected to operate on
the VLTI, should drastically reduce the jitter attenuation, hence
allowing to integrate on times much longer than the coherence time of
the atmosphere in order to reach fainter stars.


\section{The amdlib data reduction software}\label{sec_soft}
A dedicated software to reduce AMBER observations has been developed
by the AMBER consortium.  This consists in a library of C 
functions, called {\em amdlib}, plus high-level interface
programs. The {\em amdlib} functions are used at all stages of AMBER
data acquisition and reduction: in the Observation Software (OS) for wavelength
calibration and fringe acquisition, in the (quasi) Real
Time Display program used during the observations, in the online Data
Reduction Pipeline customary for ESO instruments, and in various
offline front end applications, noticeably a Yorick implementation
({\em AmmYorick}). The {\em amdlib} library is meant to incorporate all
the expertise on AMBER data reduction and calibration acquired
throughout the life of the instrument, bound to evolve with
time.

The data obtained with AMBER (``raw data'') consist in an {\em
exposure}, i.e., a time series of {\em frames}
read on the infrared camera, plus all relevant information from AMBER
sensors, observed object, VLTI setup, etc..., stored in FITS TABLE
format, according to ESO interface document VLT-ICD-ESO-15000-1826.
Saving the raw, uncalibrated data, although more space consuming,
permits to benefit afterward, by replaying the calibration sequences
and the data reduction anew, of all the improvements that could have
been deposited in {\em amdlib} in the meantime.

The library contains a set of ``software filters'' that refine the raw
data sets to obtain calibrated ``science data frames''.  This treatment
is performed on every raw data frames, irrespective of their future
use (calibration or observation). A second set of functions perform
high level data extraction on these calibrated frames, either to
compute the V2PM (see Sect.~\ref{sect:datared:p2vm}) from a set of calibration data, or to
extract the visibilities from a set of science target observations,
the end product in this case being a reduced set of visibilities per
object, stored in the optical interferometry standard $\mathrm{OI\_FITS}$ format \citep{pauls_1}.

\subsection{Detector calibration}

First, all frames pixels are tagged valid if not present in the
currently available bad pixel list of the AMBER detector. Then they
are converted to photoevent counts.  This step necessitates, for each
frame, to model precisely the {\em spatially and temporarily variable
bias} added by the electronics. The detector exhibits a pixel-to-pixel (high
frequency) bias whose pattern is constant in time but
depends on the detector integration time (DIT) and the size and
location of the subwindows read on the detector. Thus, after each
change in the detector setup, a new pixel
bias map (PBM) is measured prior to the observations by averaging
a large number of frames acquired with the detector facing a cold
shutter\footnote{Due to mechanical overheads, ``hot dark''
observations, i.e., using only an ambient temperature beam shutter
external to the dewar of the detector, are currently used to compute
the PBM.}.
This PBM is then simply removed from all frames prior to any other
treatment.

Once this fixed pattern has been removed, the detector
may still be affected by a time-variable ``line'' bias, i.e., a
variable offset for each detector line. This bias is estimated for
each scan line and each frame as the mean value of the corresponding
line of masked pixels (``DK'' column in
Fig.~\ref{fig_amber}), and substracted from the rest of the line of pixels.
The detector has an image persistence of $\sim 10\%$,
consequently all frames are corrected from this effect before
calibration.  Pixels are then converted to photoevent counts by
multiplying by the pixel's gain. Currently the map of the pixels gains
used is simply a constant  $e^-/ADU$ value (see Table~\ref{table_det})
multiplied by a ``flat field'' map acquired during laboratory tests.

Finally, the rms of the values in the
masked pixel set, that were calibrated as the rest of the detector, gives
the frame's detector noise.

\subsection{Image alignment and science data production} \label{sec_imagealign}
Once the cosmetics on the pixels is done, {\em amdlib} corrects the data
from the spatial distortions present in the image. Presently, the
only effect corrected is a displacement of the spectra acquired in the
``photometric channels'' (labeled P1, P2, P3 in Fig.~\ref{fig_amber}) with regards to the fringed spectrum in the interferometric channel. This displacement, of a
few pixels in the spectral dispersion direction, is due to
a slight misalignment of the beam-splitters described in Sect.~\ref{sec_image}, and
correcting from this effect is mandatory to compute the DC continuum
interferogram (Eq.~(\ref{eq_calculmk})). The calibration of this displacement is
performed by {\em amdlib} during the spectral calibration procedure, one of
the first calibration sequences to be performed prior to
observations. 

Finally, each frame is converted to the more handy ``science data''
structure, that contains only the calibrated image of the
``interferometric channel'' and (up to) three 1D vectors, the
corresponding instantaneous photometry of each beam, corrected from
the above mentioned spectral displacement.


\subsection{Calibration matrix computation}\label{sect:datared:p2vm}
The computation of the V2PM matrix is performed by the
function \texttt{amdlibComputeP2vm()}. This function processes the 4 or 9 files
described in Sect.~\ref{sec_calib}, applying on each of them the detector
calibration, image alignment and conversion to ``science data''
described above, then computing the $v_k^i$ (Eq.~(\ref{eq_estimFi}))
and the carrying waves $c_k^{ij}$ and $d_k^{ij}$ of the V2PM
matrix (Eq.~(\ref{eq_fitfringe})).
The result is stored in a FITS file, improperly called, for historical
reasons, ``the P2VM''\footnote{Whereas ``the V2PM'' should be the proper name.}.

The P2VM matrix is the most important set of calibration values needed
to retrieve visibilities. The shape of the carrying waves (the $c_k$s
and $d_k$s ) and in a lesser measure the associated $v_k$s, are the imprints of all the
changes in intensity and phase that the beams suffer between the
output of each fiber and the detection on the infrared camera. Any
change in the AMBER optics situated in this zone, either by moving,
e.g., a grating, or just thermal long-term effects, render the P2VM
unusable. Thus, the P2VM matrix must be recalibrated each time a new
spectral setup is called that involves changing the optical path behind the fibers.

All the instrument observing strategies and operation are governed by
the need to avoid unnecessary optical changes, and care is taken at
the operating system level to insure a recalibration of the P2VM
whenever a ``critical'' motor affecting the optical path is set in
action.
To satisfy these needs, the P2VM computation has been made mandatory
prior to science observations, and is given an unique ID number. All
the science data files produced after the P2VM file inherit of this ID,
that associates them with their ``governing'' calibration matrix. 
The {\em amdlib} library takes
the opportunity that the P2VM file is pivotal to the data
reduction, and unique, to make it a placeholder of all the other
calibration tables needed to reduce the science data, namely the spectral calibration, bad
pixels and flat field tables.

\subsection{From science data to visibilities}
The computation of visibilities is performed by the
\texttt{amdlibExtractVisibilities()} function, using a valid P2VM file.
\texttt{amdlibExtractVisibilities()} is able to perform visibility estimates on
a frame-by-frame basis, or over a group of frames, called {\em bin}.

The \texttt{amdlibExtractVisibilities()} function, in sequence:
\begin{enumerate}
\item invert the V2PM calibration matrix;
\item extract Raw Visibilities;
\item correct from Biases, compute debiased $V^2$ visibilities,
\item compute Phase Closures,
\item compute Cross Spectra
\item fit Piston values from cross spectra
\item write the OI-FITS output file.
\end{enumerate}

A typical data reduction process will first process all raw data files
related to the calibration procedures performed before acquiring the
science data, thus performing spectral calibration (e.g., using the
command line program \texttt{amdlibComputeSpectralCalibration}), then P2VM file
computation (e.g., using the command line program \texttt{amdlibComputeP2vm}).
Once the P2VM file is computed, it contains all the calibration quantities
necessary to process science object observations. One then uses the
\texttt{amdlibExtractVis} program on a science data set to get the final
OI-FITS file containing the measured science object visibilities.

\section{Illustrations and discussions}\label{sec_valid}
This section aims to present step by step the data reduction procedures performed on real interferometric measurements arising from VLTI observations. Results are discussed, focusing on key points of the process. 
\subsection{Fringe fitting} \label{sec_fringefit}
\begin{figure}[!t]
\begin{center}
\begin{tabular}{ccc}
\includegraphics[height=5cm]{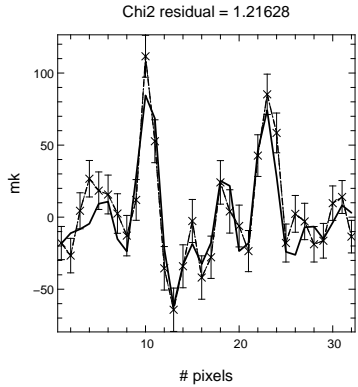} &
\end{tabular}
\caption{\label{fig_fitfringes} Example of fringe-fitting by the carrying waves, in the $3$ telescope case. The DC corrected interferogram is plotted (dashdot line) with the error bars. The result of the fit is overplotted (solid line).}
\end{center}
\end{figure}
Assuming the calibration process has been properly performed following Section \ref{sec_calib}, the first step in the derivation of the observables is to estimate the real and imaginary part of the coherent flux. This is done by inverting the calibration matrix and obtaining the P2VM matrix, shown by Eq.'s (\ref{eq_fringefit}) and (\ref{eq_invmat}). Figure \ref{fig_fitfringes} gives an example of the fringe fitting process, for an observation of the calibrator star \object{HD135382} with three telescopes. 

However, before going further in the data reduction process, it might be worthwhile for the users to check the validity of the fit and then to detect any potential problems in the data. Such step can be easily done by computing the residual $\chi^2_{res}$ between the measurements $m_k$ and the model $\widetilde{m}_k$:
\begin{equation}
[\widetilde{m}_k] = \mathrm{V2PM} [\widetilde{R}^{ij}, \widetilde{I}^{ij}]
\end{equation} 
and 
\begin{equation}
\chi^2_{res} = [\widetilde{m}_k - m_k]^T \mathcal{C}_M^{-1} [\widetilde{m}_k - m_k]
\end{equation} 
Using this checking procedure, the user can verify two critical points of the data processing:
\begin{itemize}
\item the correct subtraction of the DC component (see Eq. (\ref{eq_calculmk})): if such a condition is not fulfilled, the computed visibility will inevitably be biased since the fringe fitting by the carrying waves supposes the only presence of specific frequencies, that is the spatial coding frequencies of the instrument. A wrong DC subtraction might occur with sudden atmospheric changes between the recording of the interferometric channel and the associated photometric ones, as these channels are not on the same line of the detector, as mentioned in Section \ref{sec_imagealign}.
\item the use of a correct bad pixel map: if not, the presence of bad pixels induces high frequencies in the fringes which cannot be taken into account by the carrying waves, driving to compute biased visibility as well. Note that the bad pixel map is computed every time detector calibration is performed in the maintenance procedure.
\end{itemize} 
 \begin{figure*}[!t]
\begin{center}
\begin{tabular}{cc}
\includegraphics[height=6cm]{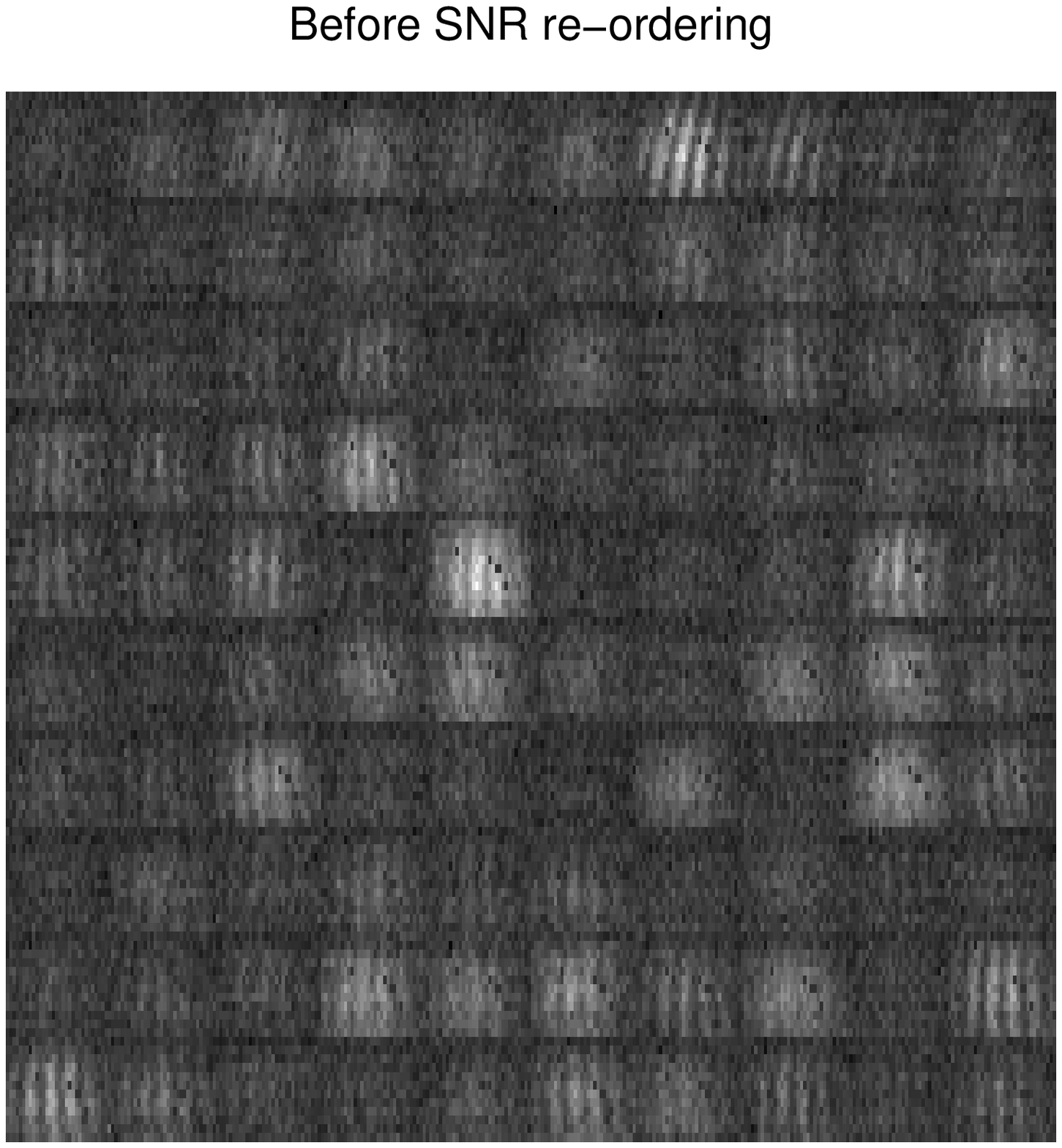} & 
\includegraphics[height=6cm]{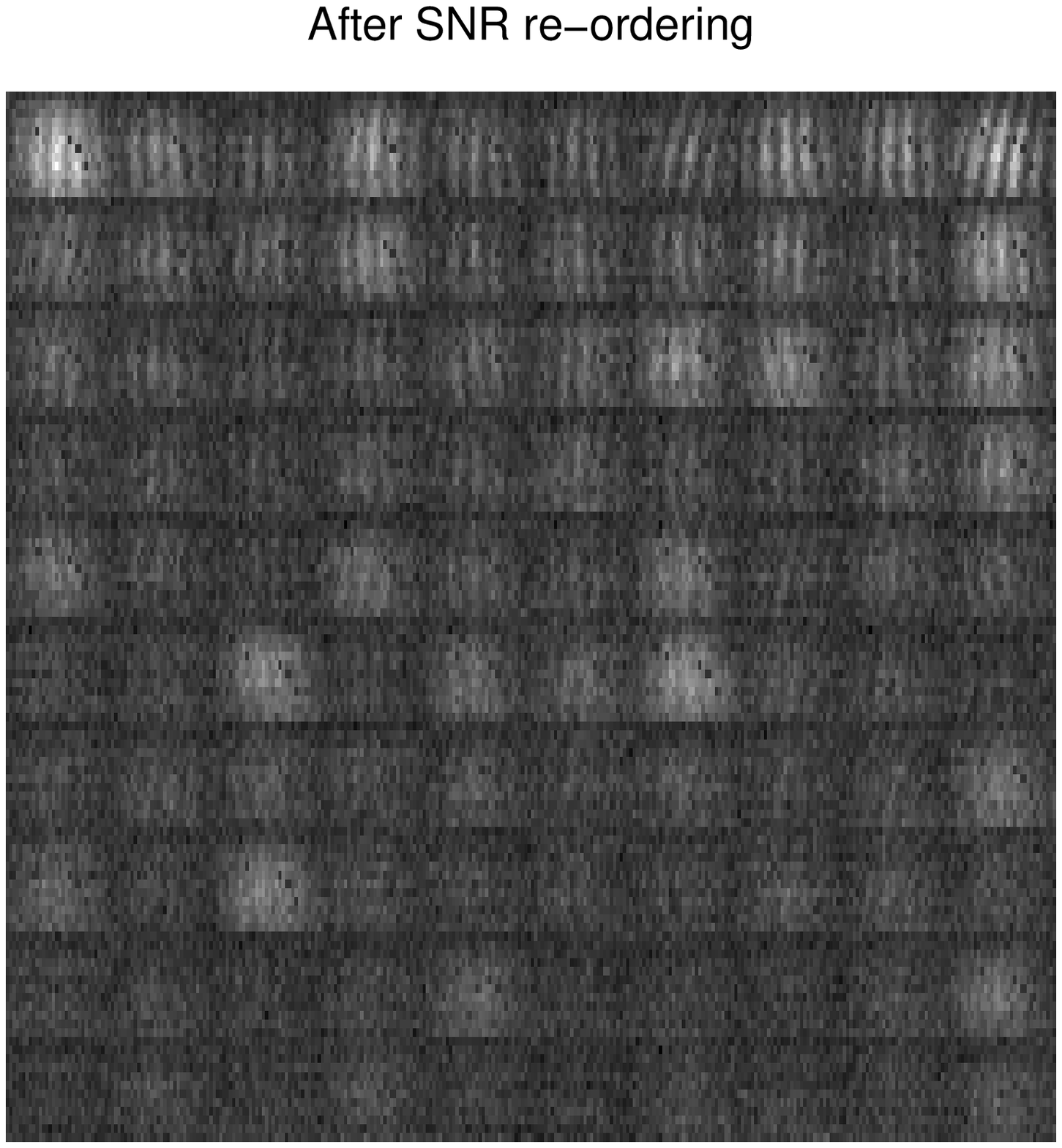} 
\end{tabular}
\caption{\label{fig_fringes} Left: sample of $100$ successive interferograms as recorded during the observation with two telescope of \object{$\epsilon~\mathrm{Sco}$} in the low spectral resolution mode. Right: Re-ordering of this sample using the fringe SNR criterion (from left to right, bottom to top). Note that, some frames which are on the bottom of the right panel (that is with relatively low SNR) appear to be brighter than some above them (that is the flux is higher). However these frames do not exhibit fringes, which explains their positions.}
\end{center}
\end{figure*}  

\subsection{Fringe criterion and fringe selection}\label{sec_fringecrit}
\begin{figure*}[!t]
\begin{center}
\begin{tabular}{ccc}
\includegraphics[height=5cm]{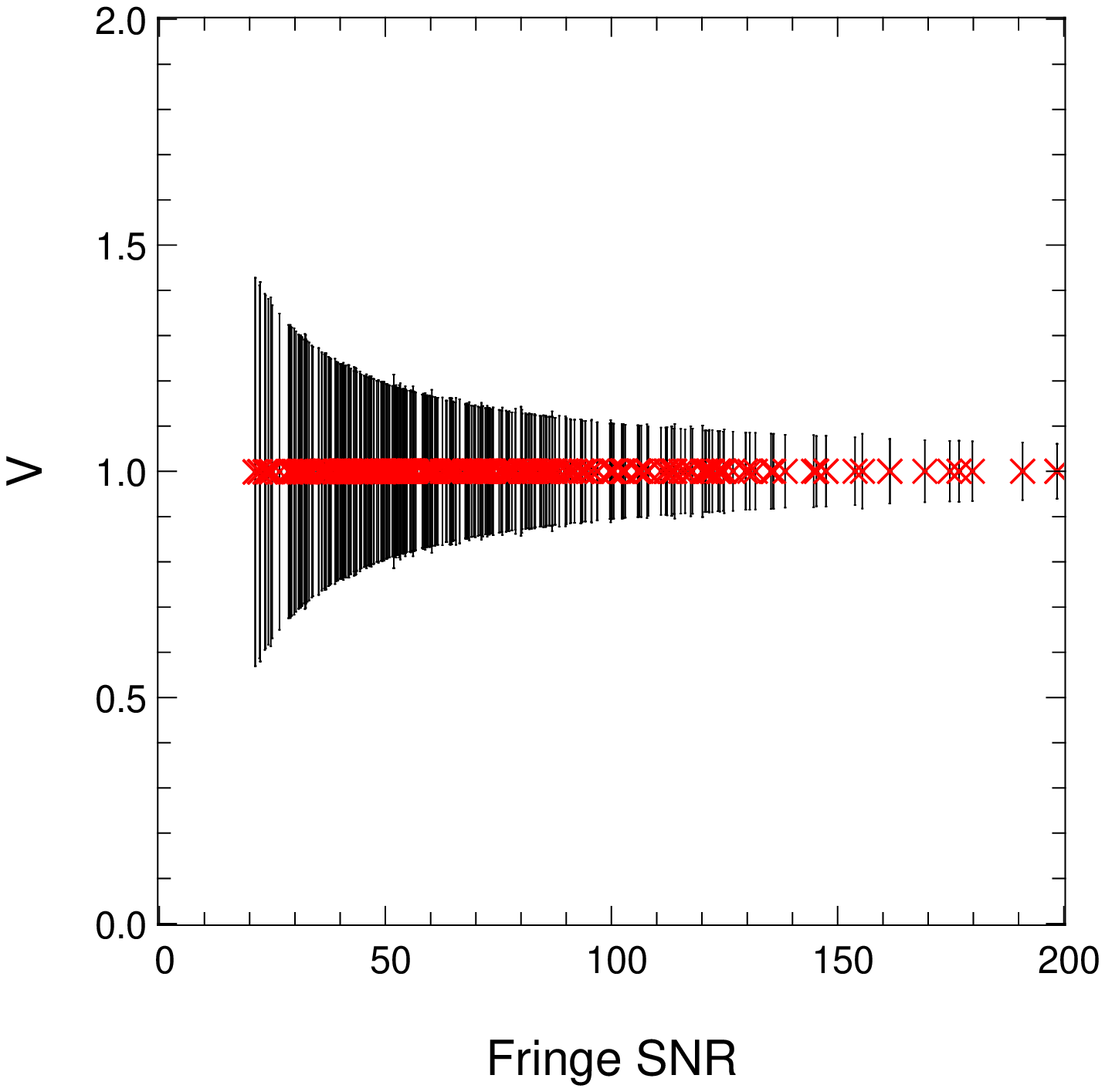} &
\includegraphics[height=5cm]{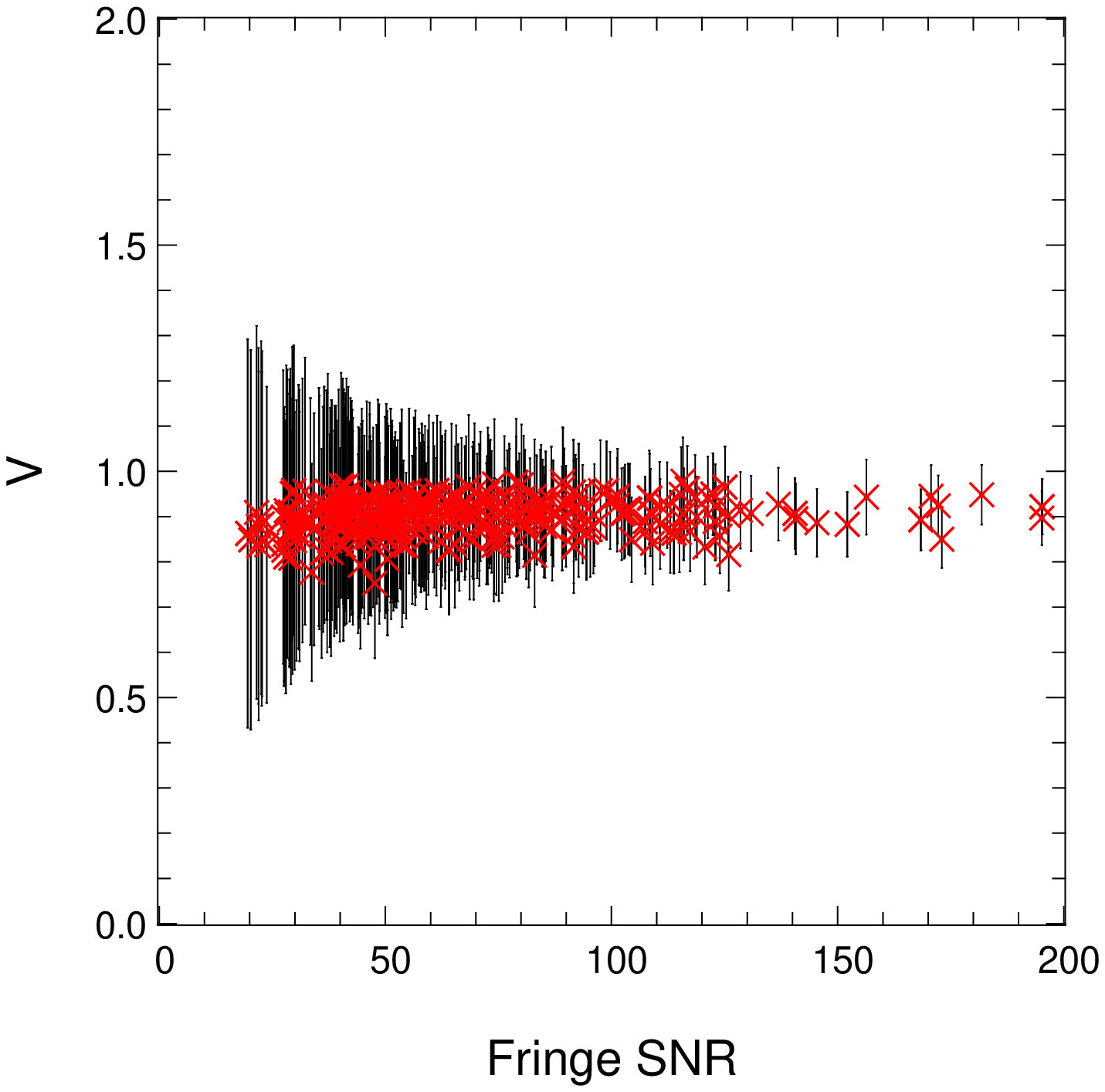} & 
\includegraphics[height=5cm]{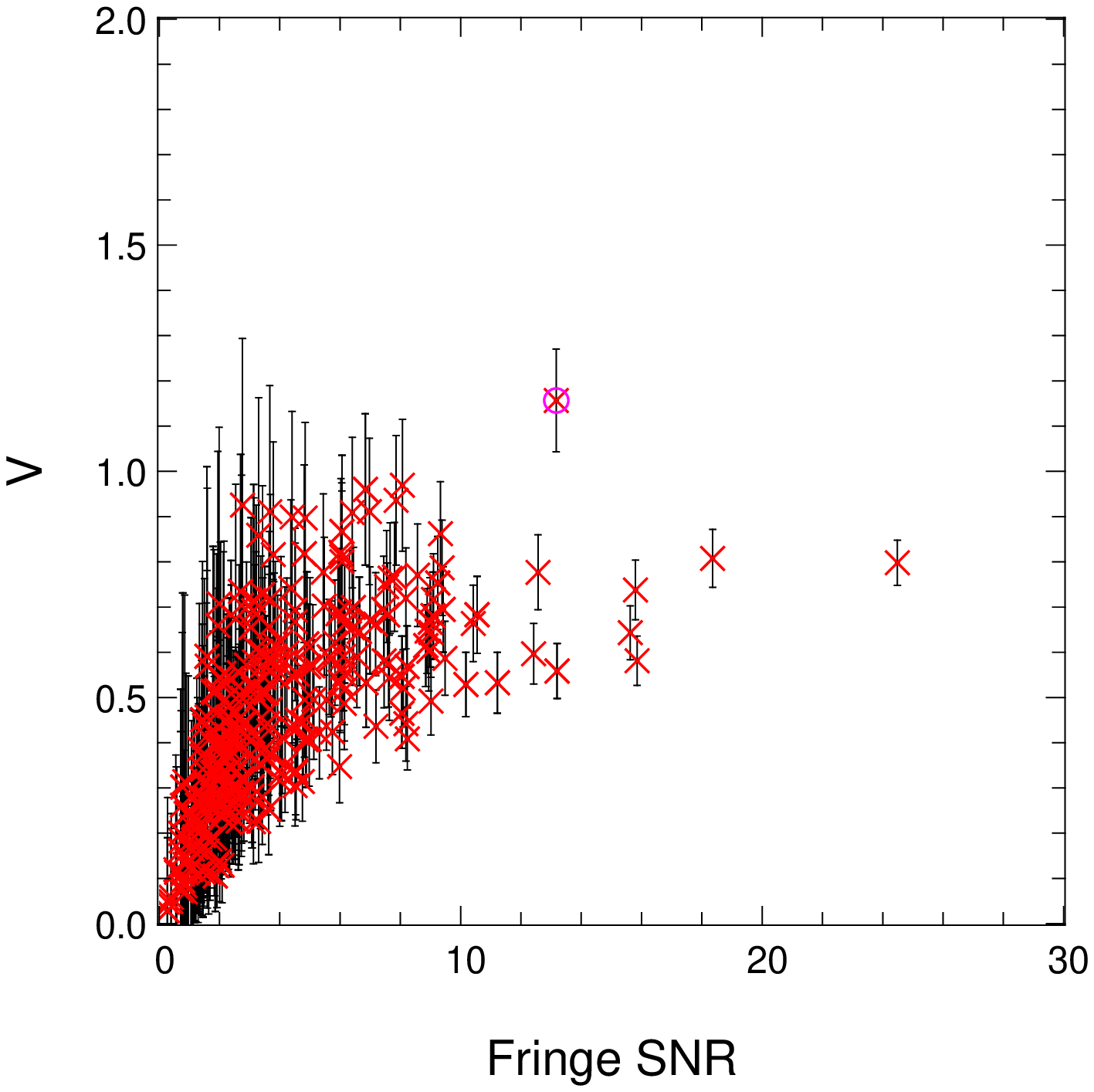} 
\end{tabular}
\caption{\label{fig_banana}Visibility as a function of the fringe SNR criterion. Left: for jitter-free simulated data, using the real photometry observed on \object{$\epsilon~\mathrm{Sco}$}. The fringe contrast was set to $1$. Middle: Same as previous one, but atmospheric jitter attenuation has been added, corresponding to a integration time of $\tau = 25\mathrm{ms}$. Right: Real \object{$\epsilon~\mathrm{Sco}$} observation. The encircled data point on the plot, well above the other ones, is typical of a bad fit of the associated fringe, as explained in Section \ref{sec_fringefit}. Note that in the first two cases, the maximum of the fringe SNR is higher than in the real case. Indeed, in the simulated data, the noise on the coherent flux only arises from the photometry $P^i$ . In the real case, however it depends as well on the noise on the interferograms $i_k$ (see Eq. (\ref{eq_sigma2mk})).}
\end{center}
\end{figure*}
For each frame of the set of data, Eq. (\ref{eq_fringeSNR}) provides an estimation of the fringe Signal to Noise Ratio.  As an example, Figure \ref{fig_fringes} presents $100$ fringes recorded on the detector during the 2-telescope observation of the calibrator \object{$\epsilon~\mathrm{Sco}$} in July 2005, first in the order as they appeared during the observation and then after re-ordering them following the fringe criterion. 

The aim of computing this criterion can be twofold: (i) during the observations, as mentioned in Section \ref{sec_fringedet}, it allows to detect the fringes and therefore to initiate the recording of the data only when it is meaningful and; (ii) calculated {\it a posteriori} during the data reduction phase, it enables to select the best frames (in terms of SNR) before estimating the observables. This second point is especially important where frames are recorded in presence of strong and variable fringe jitter. 

In the ideal and unrealistic case where the fringes are not moving during the integration time, the fringe contrast is not attenuated by vibrations, and the frame by frame estimated visibility is constant, no matter the photometric flux level in each arm of the interferometer. As a result, the visibility as a function the function of the fringe SNR is constant, with the error bars increasing as the fringe SNR decreases. This is illustrated in Figure \ref{fig_banana} (left). To obtain this set of jitter-free data, we have built interferograms  using the carrying waves of the calibration matrix that simulate perfectly stable AMBER fringes. Then, we have added the photometry taken on the \object{$\epsilon~\mathrm{Sco}$} data which allowed us to keep realistic photometric realizations taking into account the correct transmissions of the instrument. In that case, selecting the best fringes has no other ambition than improving the SNR on the observables by excluding the data with poor flux.

If the presence of atmospheric turbulence and without fringe tracker, the fringes are moving during the integration time, driving to lower the visibility. In average, the squared visibility is attenuated by a factor $ \exp(-\sigma^2_{\phi^p_{hf}})$, where $\sigma^2_{\phi^p_{hf}}$ is the variance of the atmospheric high pass jitter, as explained in Section \ref{sec_jitter}. The frame by frame visibility though, undergoes a random attenuation around this average loss of contrast. An example of the effect of the atmospheric jitter is given on Fig. \ref{fig_banana} (middle), where previous set of simulated data has been used, with adding a frame by frame random attenuation taking into account the $\tau = 25\mathrm{ms}$ integration time of the \object{$\epsilon~\mathrm{Sco}$} observation. Once again, fringe selection only enables here to increase the SNR of the observables.

However, when we look at the real set of data obtained from the observation of \object{$\epsilon~\mathrm{Sco}$}, we obtain the plot displayed on Fig. \ref{fig_banana} (right). The dispersion of the visibility, especially for low fringe SNR is unexpectedly large and can definitively not be explained by pure atmospheric OPD vibrations. As a matter of fact, these variations are due to the present strong vibrations along the VLTI instrumentation (adaptive optics, delay lines, $\ldots$), as this effect was previously revealed by the VINCI recombiner. These vibrations strongly reduce the fringe contrast and subsequently the value of the estimated visibilities, which explains the behavior of the visibilities as a function of the fringe SNR. Indeed,  when the visibility tends toward $0$, because of severe jitter attenuation, the fringe criterion tends toward $0$ as well. On the contrary, the visibility plotted as a function of the fringe SNR saturates for high values of the latter. 

The major issue is that such an effect is hardly calibratable because potentially non stationary. Hence, one convenient way to overcome the problem, beside increasing artificially the error bars to take into account this phenomena, is to only select the fringes which are the less affected by the vibrations, that is the fringes with the highest fringe SNR. One can then choose the percentage of selected frames from which the visibility will be estimated. The threshold must be chosen according to the following trade-off: reducing the number of accounted frames allows to get rid of most of the jitter attenuation, but, from a certain number -- when the sample is not large enough to perform statistics --, it increases the noise on the visibility. Furthermore, it leads to mis-estimate the quadratic bias (see Eq. (\ref{eq_biasRI2})), which is by essence a statistical quantity, and consequently drives to introduce a bias in the visibility. 

Obviously, such a selection process must be handled with care, and its robustness with regard to the selection level has to be established for any given observation. In other words, for this method to be valid, the calibrated visibility expected value must remain the same, with only the error bars changing and eventually reaching a minimum at some specific selection level. In particular, this method seems  well adapted above all to cases where the calibrator exhibits a magnitude close to the source's one, where a similar behavior of the visibility distribution versus the SNR is expected. Going in further details of this point is nevertheless beyond the scope of this paper as it will be deeply developed in \citet{millour_1}. However note that we experimentally found this procedure to be generally robust, and that for typical observations performed until now with the VLTI, choosing $20\%$ of the frames as the final sample appeared to be a good compromise.

Note that, in order to produce the curve of Fig. \ref{fig_banana}, visibilities have been computed frame by frame (i.e. $M=1$). Thus, the semi-empirical calculation of the error bars given below in Section \ref{sec_computevis} does not work, and one has to use a full theoretical expression of the noise. From an analysis in the Fourier space, \citet{petrov_2} showed that the theoretical error on the frame by frame visibility could be written:
\begin{equation}
\sigma^2(V^{ij}) = \frac{n^i+n^j+N_{pix}\sigma^2}{n^in^j}
\end{equation}
where $n^i(t)=\sum_k^{N_{pix}} v_k^i P^i(t)$ is the total flux in the $i^{th}$ beam. This computation is not fully adapted to the AMBER data processing using $3$ telescopes since in that case the Fourier peaks are overlapping. Nevertheless, it gives a rough estimation of the noise level, within a factor of $2$, which is sufficient for the analysis discussed here. 

Finally, despite that fringe selection has been performed to deal at best with the uncalibratable VLTI vibrations,  the dispersion of the selected visibilities has still to be quadratically added to the error bar arising from the fundamental noises (as computed in section \ref{sec_computevis}), in order to account for the reminiscent jitter attenuation, this latter having been reduced but not totally canceled out.

\subsection{Visibilities and associated errors}\label{sec_computevis}
\begin{figure}[!t]
\begin{center}
\begin{tabular}{ccc}
\includegraphics[height=5cm]{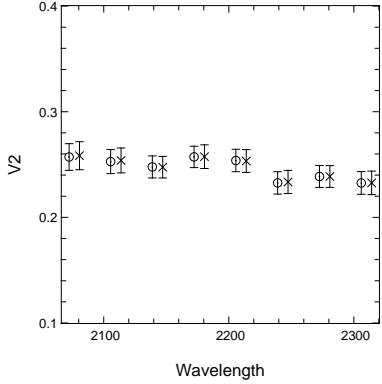} &
\end{tabular}
\caption{\label{fig_vis2}Estimation of the raw squared visibility and its error-bars as a function of the wavelength for the observed calibrator \object{$\epsilon~\mathrm{Sco}$}, in low resolution mode. Visibility crosses and corresponding errors bars are computed thanks to Eq. (\ref{eq_v2}) and Eq. (\ref{eq_sigv2_semiemp}) respectively. Circles and corresponding errors bars arise from bootstrapping technique. For sake of clarity, visibilities have been slightly shifted to the right and to the left of the corresponding wavelengths, respectively.}
\end{center}
\end{figure}
The raw squared visibility (that is biased by the atmosphere) and its associated error bar are estimated from the ensemble average of $M$ exposures, using Eq. (\ref{eq_v2}) and (\ref{eq_sigv2_semiemp}) respectively. Figure \ref{fig_vis2} gives an example of the computed squared visibility in the low resolution mode, arising from the observation of the calibrator \object{$\epsilon~\mathrm{Sco}$}. For the example considered above, we find $V^2=0.2721 \pm 0.0152$, after averaging the spectrally dispersed visibilities.   

\begin{figure}[!t]
\begin{center}
\begin{tabular}{ccc}
\includegraphics[height=5cm]{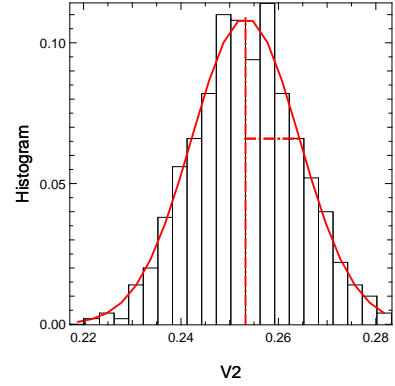} &
\end{tabular}
\caption{\label{fig_gaussfit}Histogram of the bootstrapped population of estimated squared visibilities for a given wavelength. The fit of this histogram by a Gaussian function is superimposed. The mean value and the root mean square of the Gaussian distribution gives the statistics of the estimated visibility.}
\end{center}
\end{figure}

In order to validate the computation of the error bars, we use bootstrapping techniques \citep{efron_1}. Such a method, by making sampling with replacement, constructs a large population of $N$ elements ($N$ estimated squared visibility) from the original measurements ($M$ coherent and photometric fluxes). If $N$ is large enough, the statistical parameters, that is the mean value and the dispersion of this population are converging respectively toward the expected value and the root mean square of the estimated parameters. $N$ being large enough, these quantities can be calculated by fitting a Gaussian distribution $p(V^2)$ to the histogram of the bootstrapped population. 
Figure \ref{fig_gaussfit} give an example of the histogram and the resulting Gaussian fit. Using this method with $N=500$, we find for the same set of data $V^2 = 0.2719 \pm 0.0149$, which is in excellent agreement with previous computation.

Note that, although we observed this object in the low resolution mode, that at reasonably high flux, we find a relative error of the order of $6\%$. Such a quite large error bar is due to the atmospheric and intrumental jitter that, in the absence of fringe tracking, prevents to integrate time longer than a few tenth of milliseconds. When this latter device will be we expect to lower this error below the $1\%$ level, till $0.01\%$ for the brightest cases (assuming perfect fringe tracking, see \citet{malbet_1,petrov_1}). But it is not possible to achieve AMBER's ultimate performances at that time.




\subsection{Notion of instrumental contrast in AMBER}
Given the calibration of the instrument described in Section \ref{sec_calib} and its subsequent use for the estimation of the visibility in Section \ref{eq_squaredvis}, the instrumental contrast of AMBER is self calibrated. In other words, the response of the AMBER/VLTI instrument to the observation of a point source -- in absence of atmospheric turbulence --  does not depend on the instrumental contrast but only on the visibility of the internal source (see. Eq. (\ref{eq_v2})).
Thus, if one wants to characterize the instrumental contrast, that is the total loss of contrast due to the instrumentation, one needs to use another estimator in which the calibration part (the use of the knowledge of the instrument characteristics) is skipped. We thus can use the classical definition of contrast in the image plane, directly measured ``by eyes'' from the interferograms $i_k$, recorded pair by pair of telescopes (as for the computation of the carrying waves). In spatial coding, this can be done for each pixel of the interferogram. Using Eq. (\ref{eq_interf_general}) and Eq. (\ref{eq_estimFi}), it comes:
\begin{equation}
C_r^{ij}= C_B^{ij}\left(\frac{1}{N_{pix}}\sum_k\frac{2\sqrt{P^iv^i_k P^jv^j_k}}{P^iv^i_k+P^jv^j_k}\right)
\end{equation}
Such an equation says that the instrumental contrast loss depends on two separate effects: (i) the polarization mismatch between the beams after the polarizers (vectorial effect) and; (ii) the misalignment of the interfering beams (taken into account in the product $v^i_kv^j_k$) together with the photometric unbalance between the two beams (scalar effect). Both effects are compensated when computing the visibility from the P2VM.    

\subsection{Closure phase} \label{sec_illusclosure}
\begin{figure}[!t]
\begin{center}
\includegraphics[height=5cm]{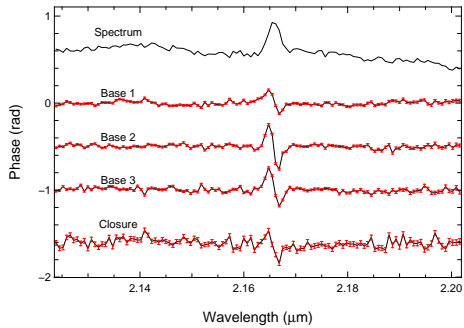}
\caption{\label{fig_phases}Example of differential phases and closure
  phase computation on an observed object with a rotating feature in
  the $\mathrm{Br}\gamma$ emission line
  (\object{$\alpha$ Arae}, see \citet{meilland_1} for a
  complete description and interpretation of these phases).}
\end{center}
\end{figure}

In the current situation, closure phases are computed using the estimator of Eq. (\ref{eq_bispectre}), but a previous frame selection is performed before making the ensemble average of the bispectrum, because in all the data available 
there was a very low amount of frames which present simultaneously three fringe
patterns. We have chosen an empirical selection criterion as the product of the three individual fringe SNR criteria (as defined by Eq. (\ref{eq_fringeSNR})).
Closure phases internal error bars are computed statistically, taking the root
mean square of all the individual frames divided by the square root of
the number of frames  (assuming statistical independence of the frames), since the tested theoretical error bars estimations does not give satisfactory
results up to now.

An example of closure phase and closure phase error bars is given in
the figure \ref{fig_phases}. The object is \object{$\alpha$ Arae} 
which contains a rotating feature in the $\mathrm{Br}\gamma$
emission line \citep{meilland_1}.
A full description on how the closure phase and closure phase errors
are computed will be part of the second paper on the AMBER data reduction \citep{millour_1}.

\subsection{Differential phase and piston}
\begin{figure*}[!t]
\begin{center}
\includegraphics[width=0.95\textwidth]{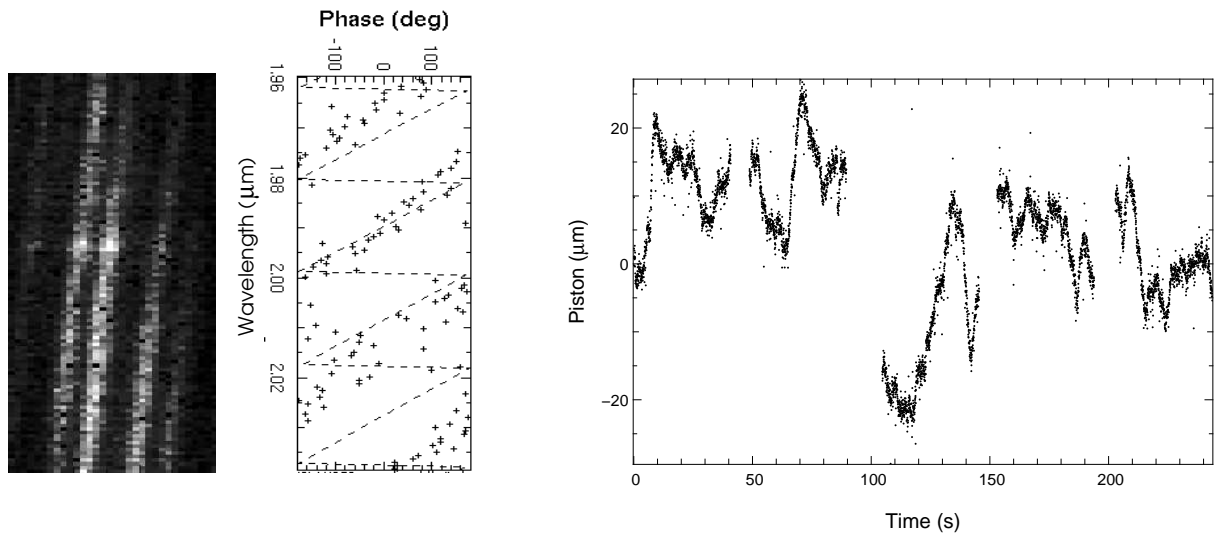}
\caption{\label{fig_piston}Piston estimation from the fringe
  pattern. From left to right is (i) the raw fringe pattern, the
  corresponding phase; (ii) the estimated linear component of the phase from the least square fit and; (iii)  a piston time-sequence over $250$ seconds. Note that the piston rms is around $15\mu\mathrm{m}$, which is in agreement with the average atmospheric conditions recorded in Paranal \citep{martin_1}.}
\end{center}
\end{figure*}
An example of differential phases is given Figure
\ref{fig_phases}. It is computed from the ensemble average of the cross spectrum as defined in the estimator of Eq. (\ref{eq_interspectre}). Frame-by-frame 
correction of its linear part (i.e. unwrapping) has been performed.
The resulting differential phase shows a typical rotation signal that
is fully described in \citet{meilland_1}.
Currently, as the closure phase, the internal error bars are computed
statistically assuming  that the differential
phases are statistically independent frame to frame. An extensive description of the data processing and of the
informations that can bring the differential phases will be described
in our second paper \citep{millour_1} .

The computation of the linear component of the differential phase, that is the piston estimation is done on each spectral band separately (J, H or
K), using the least square method described in Section \ref{sec_piston}. This
algorithm has been extensively tested on the sky and validated as a
part of the Observing Software of the AMBER instrument. An example of the fitting process as well as of the piston estimate is given in Figure \ref{fig_piston}.

\section{Conclusions}\label{sec_conc}
We have described in this paper the data reduction formalism of the VLTI/AMBER instrument, that is the principles of the algorithm that lead to the computation of the AMBER observables. This innovative signal processing is performed in three main steps: (i) the calibration of the instrument which provides the calibration matrix that gives the linear relationship between the interferogram and the complex visibility; (ii) the inversion of the calibration matrix to obtain the so-called P2VM matrix then the complex visibility and; (iii) the estimation of the AMBER observables from the complex visibility, namely the squared visibility, the closure phase and the differential phase. 

Note that this analysis requires that the calibration matrix must be both 
perfectly stable in time and very precise, that is recorded with a SNR
much higher than the SNR of the interferograms. 
If the instrument is not stable between the calibration procedures and the
observations, the P2VM will drift and as a result, the estimated
observables will be biased. And if the calibration is not precise enough, it
will be the limiting factor of the SNR of the observables. For the latter problem, it is thus recommended to set, during the calibration process, an integration time that insures a P2VM accuracy of at least a factor of $10$ higher than the accuracy expected on the measurements. To check the former problem of stability, it is advised to record one P2VM before and one P2VM directly after the observation. This procedure allows to quantify the drift of the instrument along the observations and to potentially reject the data is the drift appears to be too important. Note however that stability measurements in laboratory have shown the AMBER instrument to be generally stable at the hour scale at least.

Regarding the closure phase and the differential phase, we have produced here the theoretical estimators arising from the AMBER data reduction specific technique,  as well as brief illustrations from real observations. A thorough analysis, that is practical issues and performances, of these two observables which deal the phase of the complex visibility will be given in a forthcoming paper \citep{millour_1}

For the squared visibility, we have defined an estimator that is self-calibrated from the instrumental contrast, and we have investigated its biases. The quadratic bias, which is an additive quantity and results in the quadratic estimation in presence of zero-mean value additive noise, can be easily corrected, providing the computation of the error of the fringe measurements.  
Atmospheric and instrumental biases, which attenuate in a multiplicative manner the visibility, come respectively from the high frequency fringe motion during the integration time -- namely the jitter --, and from the loss of spectral coherence when the fringes are not centered at the zero optical path difference -- that is the atmospheric differential piston. The latter can be estimated from the differential phase and its consecutive attenuation can be corrected knowing the shape of the spectral filter and the resolution of the spectrograph. The former, when strictly arising from atmospheric turbulence, can be calibrated by a reference source, providing it has been observed shortly before/after the object of interest.  When instrumental, hardly calibratable vibrations add themselves in the jitter phenomenon, as it is presently the case for the VLTI, we propose a method based on sample selection that allows to reduce the attenuation and the associated dispersion on the visibilities.

However at this point, because of the presence of these instrumental vibrations, and because of the absence of the FINITO fringe tracker as well, it is neither possible to develop optimized tool to identify and calibrate the biases coming from the atmospheric turbulence, nor to present an analysis of the ultimate performances of the VLTI/AMBER instrument. These points will be developed in our next paper on the AMBER data reduction methods, once the problems mentioned above, which are independent of the AMBER instrument, will have been resolved.

\begin{acknowledgements}
  These observations would not have been possible without the support
  of many colleagues and funding agencies. This project has benefited
  of the funding from the French Centre National de la Recherche
  Scientifique (CNRS) through the Institut National des Sciences de
  l'Univers (INSU) and its Programmes Nationaux (ASHRA, PNPS). The
  authors from the French laboratories would like to thanks also the
  successive directors of the INSU/CNRS directors. We would like to
  thank also the staff of the European Southern Observatory who
  provided their help in the design and the commissioning of the AMBER
  instrument.
  
  This work is based on observations made with the European Southern
  Observatory telescopes. This research has also made use of
  the ASPRO observation preparation tool from the \emph{Jean-Marie
    Mariotti Center} in France, the SIMBAD database at CDS, Strasbourg
  (France) and the Smithsonian/NASA Astrophysics Data System (ADS).

  The data reduction software \texttt{amdlib} is freely available on
  the AMBER site \texttt{http://amber.obs.ujf-grenoble.fr}. It has
  been linked with the free software
  Yorick\footnote{\texttt{ftp://ftp-icf.llnl.gov/pub/Yorick}} to
  provide the user friendly interface \texttt{ammyorick}.

C. Gil acknowledges support from grant {\tt POCI/CTE-AST/55691/2004} approved by FCT and POCI, with funds from the European Community programme FEDER.

\end{acknowledgements}


\end{document}